\documentclass[preprint,showpacs,preprintnumbers,amsmath,amssymb, superscriptaddress,longbibliography,nofootinbib]{revtex4}

\usepackage{textcomp}
\usepackage{makeidx}
\usepackage{amsmath}
\usepackage{subfigure}
\usepackage{amssymb}
\usepackage{hyperref}
\usepackage{graphicx}
\usepackage{epstopdf}
\usepackage{color}
\usepackage{soul}
\usepackage[T1]{fontenc}

\begin{document}

\title{Regular black holes with $\Lambda>0$ and its evolution in Lovelock gravity.}

\author{Rodrigo Aros}
\email{raros@unab.cl}
\affiliation{Departamento de Ciencias Fisicas, Universidad Andres Bello, Av. Republica 252, Santiago,Chile}

\author{Milko Estrada}
\email{milko.estrada@gmail.com}
\affiliation{Department of Physics, Universidad de Antofagasta, 1240000 Antofagasta, Chile}

\date{\today}

\begin{abstract}
In this work it is shown that the thermodynamics of regular black holes with a cosmological horizon, which are solutions of Lovelock gravity, determines that they must evolve either into a state where the black hole and cosmological horizons have reached thermal equilibrium or into an extreme black hole geometry where the black hole and cosmological horizons have merged. This differs from the behavior of Schwarzschild de Sitter geometry which evolves into a de Sitter space, the ground state of the space of solutions. This occurs due to a phase transition of the {\it heat capacity} of the black hole horizon. To perform that analysis it is shown that at each horizon a local first law of thermodynamics can be obtained from the gravitational equations.     
\end{abstract}


\maketitle

\section{Introduction}

Motivated by the many known analogies between the physics of black holes and thermodynamics, there has been a great interest to analyze how deep these relations hold and how much they can be extended  \cite{Bardeen:1973gs,Bekenstein:1973mi,Bekenstein:1973ur,Hawking:1974sw}. Indeed, nowadays there are several research programs exploring the idea that gravity may be an emergent theory with thermodynamic roots. In reference \cite{Cosmo} was shown that the Friedmann equations can also be interpreted as thermodynamic identities. Certainly, one can check that this is well studied topic in the literature \cite{Aplicacion4,Aplicacion5,Aplicacion6,Aplicacion7,Aplicacion8,Aplicacion9}. For this work, however, it is more relevant to mention that the connection between the gravitational field equations at the horizons and the first law of thermodynamics. For instance, in \cite{Pad1} was shown that the Einstein equations in the presence of an anisotropic fluid (with a mass density $\rho$ and a radial pressure $P=-\rho$) on any horizon can be shaped up into the form of a first law of thermodynamics $PdV=TdS-dU$. The temperature $T$ is defined as the inverse of the Euclidean period\footnote{This coincides with the thermal definition in \cite{Hawking:1974sw}} at the horizon. In \cite{Pad2} this was generalized to the Kerr-Newman solution and a time dependent horizons. Two different questions arise after this. First, what does happen when the geometry has more than a single horizon and thus there is more than a single temperature?. Considering thermodynamics the answer is that this must represent a non-equilibrium system and therefore a system that must evolve consequently \cite{Aros:2008ef}. The second question is, how are connected the local thermodynamics relations defined in \cite{Pad1}, with the standard black hole thermodynamic relations which are essentially non-local? This work aims to shed some light on those problems. . 

\subsection*{Schwarzschild de Sitter space and thermodynamics}

Originally, the first law of thermodynamics of black holes was discovered by analyzing black hole geometries with a well defined asymptotic region. Roughly speaking, this law can be understood as the relation between the changes of the mass, angular momenta and electric charges, which are to be computed at the asymptotic region, and the changes at the horizon. In fact, the first law was re-derived in terms of the analysis of the variation of Noether's charges in \cite{Wald:1993nt}.  

From the above, it seems that spaces with cosmological horizons may challenge the existence of a first law of thermodynamics since they have no asymptotic region and consequently conserved charges cannot be defined in the usual way. Fortunately, this is not the case. Following \cite{Lee:1990nz}, for instance, the first law of the thermodynamics turns into a sort of balance equation of the heat flows between the two horizons. This is complete consistent with the picture of a non-equilibrium system mentioned above. This can be visualized by considering the Schwarzschild-dS solution in $d$-dimensions where the derivation \`a la Wald \cite{Wald:1993nt}, with the black hole and cosmological horizons defined by $r_+$ and  $r_{++}$ respectively, yields \cite{Aros:2008ef} 
\begin{equation}\label{ConservationofHeat}
     T_{+} dS_{+} = -T_{++} dS_{++},
\end{equation}
where $T_+$ and $T_{++}$ are the temperatures at the black hole and cosmological horizons respectively. $dS_{+}$ and $dS_{++}$ are the variations of the {\it entropies/areas} of both horizons. Indeed, the conservation can be understood as $dQ_{+} = -d Q_{++}$. 

In this moment one may question what is the role of the local thermodynamic relations at each horizon of the proposal in \cite{Pad1}. Indeed, the existence of a local and non local first laws of thermodynamics may seem contradictory but actually it is not. For Schwarzschild-dS solution, with a mass parameter $M$, the local relations are respectively $dM =  T_{+} dS_{+}$  and $dM = -T_{++} dS_{++}$, and thus  Eq.(\ref{ConservationofHeat}) is nothing but the combination both local thermodynamic relations. It must be stressed that the scenario in the presence of an (anisotropic) fluid is not that simple. It is worth to mention \cite{Dymnikova1} where the local approach was applied to multiple horizon geometries. 

\subsection*{Lovelock Gravity}
In higher dimensions ($d>4$) Lovelock gravities arise as a simple generalization of the Einstein gravity. For instance, Lovelock gravity's equations of motion are of second order. The Lovelock gravity Lagrangian is given by
\begin{equation}
    L = \sum_{p=1}^{[d/2]} \alpha_p L_p.
\end{equation}
where $\{\alpha_p\}$ is a set of arbitrary coupling constants and $L_p$ is conventionally normalized such as
\begin{equation}\label{Ln}
L_p = \frac{1}{2^p} (d-2p)!\delta^{\mu_1 \ldots \mu_{2p}}_{\nu_1 \ldots \nu_{2p}} R^{\nu_{1} \nu_{2}}_{\hspace{2ex}\mu_{1} \mu_{2}}\ldots R^{\nu_{2p-1} \nu_{2p}}_{\hspace{2ex}\mu_{2p-1} \mu_{2p}},
\end{equation}
where $R^{\alpha \beta}_{\hspace{2ex}\mu\nu}$ is the Riemann tensor and $\delta^{\mu_1\ldots \mu_n}_{\nu_1 \ldots  \nu_n}$ is the generalized $n$-antisymmetric Kronecker delta.

In principle, the Lovelock gravity equations of motion may have as many different constant curvature solutions as the highest power of the Riemann tensor presented in the Lagrangian \cite{Edelstein,Arenas-Henriquez:2017xnr}. This defines a set of effective cosmological constants $[ \Lambda_1,\ldots \Lambda_n ]$. These constant curvature solutions are to be identified with the ground states of the Lovelock gravity considered. 

In \cite{milko} was studied the asymptocally locally AdS regular black hole geometries with a unique AdS ground state. In this work it will be studied the case with a unique de Sitter ground state. There are two cases of interest. The first one is when the equation of motion have, roughly speaking, the form $(R-\Lambda)^{n} = 0$. In this case there is a single, but $n$-fold degenerated, ground state of curvature $\Lambda>0$. This can be considered a {\it kinetic} selection of the ground state\footnote{This case is analogous to the negative cosmological constant case analyzed in \cite{Crisostomo:2000bb}}. The second case is when only one of the effective cosmological constant is a real positive number. For simplicity, the case of interest in this work is when the  Lagrangian is just a single term in the Lovelock series plus a cosmological constant, {\it i.e.},  $L = \alpha_n L_n + \alpha_0 L_0$. This theory is known as {\it Pure Lovelock}. See for instance \cite{Cai:2006pq, Toledo:2019mlz,Toledo:2019szg, Dadhich:2015ivt}. This theory, for $\Lambda>0$, has a unique de Sitter ground state $\forall n$. For $n$ even, however, there is an additional anti de Sitter ground state which is not relevant for this work, and thus it will be ignored.

 Finally, it is worth to mention references \cite{Aplicacion1,Aplicacion2,Aplicacion3,Pad3} where the connection between generic Lovelock gravity and the local thermodynamic relation at the horizons were studied. 

\subsection*{Thermodynamic evolution}

The thermodynamics of a solution with a de Sitter ground state is expected to differ from those which are asymptotically flat or AdS in many aspects. The most remarkable difference is due to an effective positive cosmological constant which usually implies the presence of a cosmological horizon, and  thus the lack of an asymptotic region. 

Now, as mentioned above, if a geometry has a black hole and a cosmological horizon, this can naturally be interpreted as non equilibrium thermodynamic system \cite{Aros:2008ef,Gomberoff:2003ea} and therefore a system that must evolve. The evolution of the vacuum solutions of the $n$-fold Lovelock gravity with a de Sitter ground state was studied in \cite{Aros:2008ef}. In that case the thermodynamics determines that geometry evolves from a two-horizons geometry into the de Sitter space, the ground state of the theory. This happens due to the temperature of the black hole horizon is always larger than the cosmological horizon one, while the heat capacities, defined as are always negative for the black hole horizon and positive for the cosmological horizon. By treating this system as any ordinary thermodynamic system, this implies that while the cosmological horizon aims to increase its temperature due to the hotter radiation it absorbs. The black hole horizon, on the other hand, aims to increase its temperature as well due to the negative heat capacity. The only compatible scenario is therefore that $M$ must decrease during the process. This process will continue as long $M \neq 0$. The whole evolution can be interpreted as a process in which black horizon emits more and more radiation until eventually $M$ is entirely radiated away into the cosmological horizon. Conversely, the analysis of the Pure Lovelock \textbf{vacuum} solutions shows that the temperature of the cosmological horizon in fact can be larger than the black hole one (within a region of the space of parameters). This has a profound impact in the evolution of the Pure Lovelock solutions as will be discussed below. The analysis of a  a generic Lovelock theory in this respect will be presented in a future work. 

\section{Evolution of regular Black Holes}

To begin with the analysis it is worth to recall that regular black hole solutions emerge from the idea that corrections, due to \emph{would-be} quantum gravity, should give rise to the removal of the gravitational singularities presented in standard black hole geometries. The point is that those corrections can be mimicked by the introduction an anisotropic fluid which must satisfy among other conditions, see \cite{milko} for a discussion, to be strongly\ concentrated at the origin. For simplicity in this work it will be described only the static case. In Schwarzschild coordinates,
\begin{equation} \label{elementodelinea}
ds^2 =-f(r) dt^2+ \frac{dr^2}{f(r)} + r^2 d \Omega^2_{D-2},
\end{equation}
the energy momentum tensor of any fluid must be given by
\begin{equation}\label{TFluid}
    T^{\alpha}_{\hspace{1ex}\beta} =  \textrm{diag}(-\rho, p_r, p_\theta, p_\theta, ...),
\end{equation}
where $\rho=-p_r$. Moreover, due to the conservation law $\nabla_{\mu} T^{\mu\nu} =0$,
\begin{equation}
p_\theta = \frac{r}{d-2} \frac{d}{dr}p_r + p_r.
\end{equation}
This determines that this is an {\it anisotropic fluid}. Before to proceed it is convenient to define the mass function
\begin{equation}\label{massfunction}
  m(r) = \Omega_{d-2} \int_0^r \rho(\xi) \xi^{d-2} d\xi,
\end{equation}
where $\Omega_{d-2}$ is the {\it volume} of the $d-2$ dimensional transverse section. In order to mimic the quantum effects the mass density $\rho$ must have a singles maximum at $r=0$ and rapidly to go to zero with $r$. This implies that \cite{milko}
    \begin{equation} \label{Masacercaorigen}
        \left. m(r) \right|_{r \approx 0} \approx C_1 r^{d-1}
    \end{equation}
with $C_1$ a positive constant. This also implies that the geometry around $r=0$ is a de Sitter space. 

For spaces with an asymptotic region, {\it i.e.} for spaces where $r \rightarrow \infty$ is a proper limit, $m(r)$ must tend to a constant $M$ which is proportional to the mass of the solution \cite{milko}. Now, although for spaces with a cosmological horizon the mass of a solution is not well defined, still it is necessary that the solution have a {\it mass parameter} $M$, whose vanishing defines the ground state, {\it i.e.}, a de Sitter space.  By the same token, $m(r)$ must be also promoted to a function $m(r,M)$ such as $m(r,M)|_{M=0} = 0$. Moreover, the mass function $m(r,M)$ must increase with the mass parameter. Therefore, 
\begin{equation}
 \frac{\partial m(r,M)}{\partial M} > 0,
\end{equation}
for any value of $r$. $M$ will be called from now on the mass parameter of the solution

The definitions above imply that at any horizon, says $r=a$ with $f(a)=0$, arises an effective mass function $m(a,M)$. Now, although $m(r,M)|_{r=a} = m(a,M)$ defines the value of the mass density at the horizon, once evaluated $m(M,a)$ is to be understood as function only of the horizon radius, $m(a)$. This will be relevant in the next sections and shown explicitly. 

In the next sections will be shown that the general case has three horizons, defined by the radii $r_{-} < r_{+} < r_{++}$. As mentioned above, the construction of a non-local thermodynamics from the local thermodynamic is not direct in the presence of fluid, but still the analysis can be carried out to a certain degree thanks to the presence of the mass parameter $M$. First, it must be noticed that to perform the analysis $f(r)$ must be formally promoted to $f(r,M)$. Once this is established one can proceed. Let $r=a$ be a zero of $f(r,M)$. Now, in order to map a solution whose horizon is defined by $r=a$ with a mass parameter $M$ into a solution with a horizon at $r=a+da$ then one must consider the transformation in the space of solutions $a \rightarrow a  + da$ and $M \rightarrow M + dM$ such as $f(a+da,M+dM) = 0$ or equivalently 
\begin{equation}\label{evolutionBH}
    df(a,M) = 0 = \frac{\partial}{\partial a} f(a,M) da + \frac{\partial}{\partial M} f(a,M) dM.
\end{equation}
This defines, in turn, the relation a long the curve of solutions,
\begin{equation}
    dM = - \left( \frac{\partial}{\partial M} f(a,M) \right)^{-1}  \left( \frac{\partial}{\partial a} f(a,M) \right) da.
\end{equation}

It must be noticed that the definition of the temperature of a cosmological horizon requires certain considerations due to the orientation of the normal vectors. While the radial derivative is outward to the black hole horizon, it is inward to the cosmological horizon. This defines the direction of the period clockwise for the cosmological horizon. Because of that it is necessary to define the temperatures as 
\begin{eqnarray}
      T_{+}  &=&  \frac{1}{4\pi} \left. \frac{\partial f(r)}{\partial r} \right|_{r=r_{+}} \textrm{ but}\\
      T_{++} &=& - \frac{1}{4 \pi} \left. \frac{\partial f(r)}{\partial r} \right|_{r=r_{++}}.
\end{eqnarray}
With this convention the both temperatures are positive. See for instance \cite{Nam,Dymnikova1}. Under this convention one obtains the non-local relation
\begin{equation} \label{RelacionHorizontes}
 dM=T_+ \left( \frac{\partial}{\partial M} f(r_+,M) \right)^{-1} dr_{+}  = - T_{++} \left( \frac{\partial}{\partial M} f(r_{++},M) \right)^{-1} dr_{++}.
\end{equation}
This relation will be made explicit for the two Lovelock theories considered in the next sections. 

In order to study the thermodynamics as for any other standard thermodynamic system one can define a heat capacity. In this case, however, there are two potential generalizations. First, one can define \cite{Bunster1,AplicacionDS3}
\begin{equation}\label{HEATCAPACITY1}
K(a) =  sign \left. \frac{\partial M}{\partial T} \right|_{r=a}  = -4\pi T(a) \left( \frac{\partial}{\partial M} f(a,M) \right)^{-1}  \left( \frac{\partial }{\partial a} T \right)^{-1} ,
\end{equation}
where $sign = 1$ for $r=r_+$ the black hole horizon. Analogously, $sign = -1$ for $r=r_{++}$, cosmological horizon. This is a direct generalization of the definition of the heat capacity for vacuum (black hole) solutions. It must be noticed that although this is a local definition for each horizon, due to the mass parameter the changes are not independent and thus this provides with an element to analyze the evolution. For the vacuum solutions the mass parameter coincides with a local definition of energy $M=U_{a=r_+}=-U_{a=r_{++}}$ \cite{Bunster1,Gomberoff:2003ea}.  

The second defintion of a heat capacity comes from the local thermodynamic relations proposed in \cite{Pad1}, in the form of
\begin{equation}\label{heatcapacity2}
    C(a) = sign \left.\frac{\partial m(r,M)}{\partial T} \right|_{r=a}.
\end{equation}
Unlike the conventional definition of heat capacity, the equation (\ref{heatcapacity2}) can not represent the usual $(\partial Q/ \partial T)_{V}$ or $(\partial Q/ \partial T)_{S}$ as the change of volume and entropy are both functions of horizon's radii, and thus not mutually independent. However, as we will see below, Eq.(\ref{heatcapacity2}) will be indeed useful to analyze the evolution of the solution. On the other hand, it is straightforward to notice that Eq.(\ref{heatcapacity2}) differs from Eq.(\ref{HEATCAPACITY1}) in some formal aspects. In the the next sections, however, when equations (\ref{HEATCAPACITY1}) and (\ref{heatcapacity2}) will be analyzed for some particular examples, it will be noticed that both provide essentially the same information about the evolution.

\section{Pure Lovelock Regular Black Holes}
Pure Lovelock theory, as mentioned above, is defined by the Lagrangian $L_n + L_0$ and its black hole solutions in vacuum  have been studied. See \cite{indio1,indio2,Cai:2006pq} for instance. In \cite{Cai:2006pq} are discussed the thermodynamics features of the vacuum solution with a de Sitter ground state.

For $\Lambda>0$ the equations of motion for a regular black hole were discussed originally in \cite{milko}. The gravitational equations are actually just reduced to solve
\begin{equation} \label{eqmotionPL}
    \rho r^{d-2} \Omega_{d-2} + \frac{d-1}{l^{2n}}r^{d-2} = \frac{d}{dr} \left( r^{d-2n-1} (1-f(r,M))^n \right),
\end{equation}
where $\Omega_{d-2}$ stands for volume of the transverse section. By direct integration,
\begin{equation} \label{solucionPL}
    f(r,M)= 1-\left( \frac{m(r,M)}{r^{d-2n-1}}+ \left(\frac{r^2}{l^2}\right)^n \right)^{1/n},
\end{equation}
where $d-2n-1>0$ and $m(r)$ is the mass function defined in Eq.(\ref{massfunction}).

\subsection{Structure of the horizons}
\begin{figure}
\centering
{\includegraphics[width=4in]{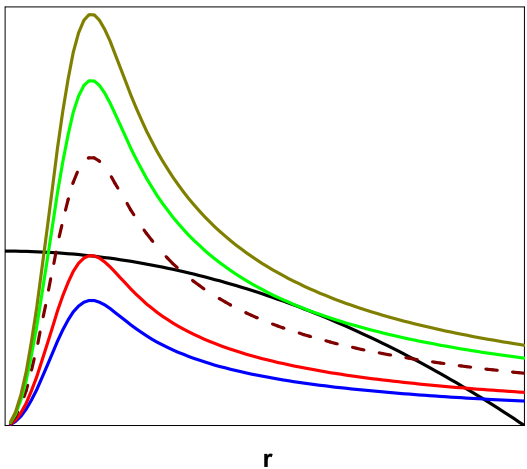}}
\caption{zeros of $f(r)$}
\label{zerostructure}
\end{figure}

Although, no analytic expression for $m(r,M)$ has been given yet, still it is possible to analyze the structure of zeros of $f(r,M)$. This can be done by drawing the functions
\begin{equation} \label{zero1}
\left( 1-\frac{r^{2n}}{l^{2n}}, \frac{m(r,M)}{r^{d-2n-1}} \right),
\end{equation}
whose intersection defines the zeros of $f(r,M)$. In the Fig. (\ref{zerostructure}), the black curve represents $1 - (r/l)^{2n}$. The colored set of curves corresponds to $m(r,M) r^{1+2n-d}$, depicted for a particular $m(r,M)$ with increasing values the mass parameter. Recall that $\partial m/ \partial M >0$. 

Although, for this picture one particular $m(r,M)$ has been chosen, it is straightforward to realize that the behavior must be generic, as along as $\rho$ satisfies the criteria mentioned in \cite{milko}. Therefore, there are up to three zeros of $f(r,M)$ for an arbitrary function $m(r,M)$. These are to be called  $r_{-}<r_{+}<r_{++}$. The outer zero defines the presence of a cosmological horizon $r_{++}$, and thus defines the outer spatial boundary of the space. $r_-$ and $r_+$, represent an internal and black hole horizon, respectively. 

Moreover, one can notice that there are three horizons only within a finite range of the mass parameter, says defined by $M \in [M_{cri1},M_{cr2}]$. $M=M_{cri1}$ defines when $r_{+}$ and $r_{-}$ merge. Analogously, $M=M_{cri2}$ defines when $r_+$ and $r_{++}$ merge and is known as {\it Nirai space time} \cite{Nirai}. In the figure (\ref{zerostructure}) , the curves are displayed with color blue (for $M<M_{cri1}$), red (for $M=M_{cri1}$), wine (for $M_{cri1}<M<M_{cri2}$), green (for $M=M_{cri2}$) and dark yellow (for $M>M_{cri2}$).

Due to the absence of singularities of regular black holes geometries, indeed they are not limited by any {\it cosmological censorship theorem}, it seems tempting to explore solutions with three, two and one horizons. They are mathematically sound. However, that does not make them physically relevant in the context of this work. It must be noticed that although for both $M< M_{cri1}$ and $M> M_{cri2}$ there is a single horizon geometry, they correspond to completely different physical scenarios. $M< M_{cri1}$ can be only understood as a star, due to the lack of a black hole horizon. Notice that in this case no singularity can arise at $r=0$. Conversely, $M>M_{cri2}$ can be understood as mass density that extended to cosmological scales. This case is considered nonphysical even for the vacuum solutions. These two scenarios are not physical relevant for this work.  

\subsection{Thermodynamics redone}
In order to analyze the thermodynamics of these spaces, $r=a$ will denote a generic horizon subjected, for now, only to satisfy $f(a)=0$. The direct manipulation of equation (\ref{solucionPL}) yields
\begin{equation}
    m(M,a)= a^{d-2n-1}- \frac{a^{d-1}}{l^{2n}}.
\end{equation}
One can noticed that although $m(M,r)$ is a function of $M$ and $r$, once evaluated on a horizon becomes just a function of the horizon radius, in this case $a$. 
Following ref. \cite{Pad2}, by evaluating the equation of motion (\ref{eqmotionPL}) at $r=a$, identifying $\rho=-P$, yields,
\begin{equation}
    P d \left( \frac{\Omega_{d-2}a^{d-1}}{d-1} \right) = \frac{f'}{4\pi} d \left( 4 \pi \frac{n a^{d-2n}}{d-2n} \right) - d \left( a^{d-2n-1}-\frac{a^{d-1}}{l^{2n}} \right)
\end{equation}
or equivalently
\begin{equation} \label{primeraley}
    P d \left( \frac{\Omega_{d-2}a^{d-1}}{d-1} \right) = \frac{f'}{4\pi} d \left( 4 \pi \frac{n a^{d-2n}}{d-2n} \right) - d \left( m(a) \right).
\end{equation}
This relation can be understood as a constraint in the evolution of the solution, within the space of parameters, along the change $a$ and $a+da$. As mentioned above, the same relation can be obtained by requiring that $f(a+da) =0$ along the transformation $a \rightarrow a + da$.

The interpretation of Eq.(\ref{primeraley}) as a local thermodynamics at each horizon arises directly \cite{Pad1,Cai:2006pq, Bunster1}. For $a=r_+$ one identifies, 
\begin{align}
    \mbox{Thermodynamics Pressure}&=P(r=r_+) \label{presionBH}\\
                \mbox{Volume ($V$)}&=\frac{\Omega_{d-2}r_+^{d-1}}{d-1} \label{volumenBH}\\
\mbox{Temperature BH horizon ($T_+$)}&=\frac{f'}{4\pi} |_{r=r_+} \label{TemperaturaBH}\\
\mbox{Entropy ($S_+$)}&= 4 \pi \frac{n r_+^{d-2n}}{d-2n}, \\
\mbox{Energy BH horizon ($U_+$)}&=m(r_+). \label{energiaBH}
\end{align}
Notice that the $f'|_{r=r_+}>0$ is satisfied strictly. In addition, $m(r_+)$ can be interpreted as a local notion of energy. With this in mind, the equation takes the thermodynamic form 
\[ P_+dV_+=T_+dS_+-dU_+.
\] 
Analogously, If $a=r_{++}$ one needs to identify
\begin{align}
    \mbox{Thermodynamics Pressure}&=P(r=r_{++}) \label{presionCH} \\
                \mbox{Volume ($V$)}&=\frac{\Omega_{d-2}r_{++}^{d-1}}{d-1} \label{volumenCH} \\
\mbox{Temperature cosmological horizon ($T_{++}$)}&=-\frac{f'}{4\pi}|_{r=r_{++}} \label{TemperaturaCH} \\
\mbox{Entropy ($S_{++}$)}&= 4 \pi \frac{n r_{++}^{d-2n}}{d-2n}, \\,
\mbox{Energy cosmological horizon ($U_{++}$)}&=-m(r_{++}), \label{energiaCH}
\end{align}
to obtain $P(-dV)=TdS-dU$. Notice that $f'|_{r=r_{++}}<0$. On the other hand, $P(-dV)$ has the correct sign since the inaccessible region (where $f<0$) is now outside the cosmological horizon and thus the volume of the outer region change as $-dV$ if $r_{++} \rightarrow r_{++} + dr_{++}$ with $ dr_{++} > 0$ \cite{Pad1}. 

Finally, the heat capacities $ C_{+}$ and $C_{++}$ mentioned above in equation (\ref{heatcapacity2}) in this case are
\begin{align}
    &C_+=\frac{dU(r_+)}{dT}=\frac{dm(r_+)}{dT} =  r_+^{d-2n-1}\left( \frac{d-2n-1}{r_+} - \frac{r_+^{2n-1}}{l^{2n}} \right) \left(\frac{\partial T_+}{\partial r_+} \right)^{-1}\label{C1aley1} \\
    &C_{++}=\frac{dU(r_{++})}{dT}=-\frac{dm(r_{++})}{dT} =   r_{++}^{d-2n-1}\left( -\frac{d-2n-1}{r_{++}} + \frac{r_{++}^{2n-1}}{l^{2n}} \right) \left(\frac{\partial T_{++}}{\partial r_{++}} \right)^{-1}. \label{C1aley2}
\end{align}
One can notice that the second factor of in Eqs. (\ref{C1aley1}) and (\ref{C1aley2}) is expression of the temperature of the vacuum solution. It is straightforward to demonstrate that the second factor in both expression, for $r_+$ and $r_{++}$, is positive. Therefore, the signs of the specific heats $C_+$ and $C_{++}$ depend only on the sign of $\partial T/ \partial a$. 

On other hand, from equation (\ref{HEATCAPACITY1}), following the convention in \cite{AplicacionDS3}, the $K$ heat capacity of the black hole horizon ($r_+$) is given by
\begin{equation}\label{PureLovelockKexplicit}
  K_{+} = 4 \pi n r_{+}^{d-2n-1} \cdot T_+ \cdot \left.\left( \frac{\partial m(r,M)}{\partial M} \right )^{-1}\right|_{r=r_+} \cdot  \left(\frac{\partial T_{+}}{\partial r_{+}} \right)^{-1}.
\end{equation}
The expression for $r_{++}$ is analogous. Obviously, an explicit expression for $m(r,M)$ is required to obtain the exact expressions, however, it is direct to see that the three first factors are always positive for $a=r_+$ and $a=r_{++}$. Therefore, the sign of the specific heat depends only on the sign of the factor $\partial T/ \partial a$ as previously for Eq.(\ref{C1aley1},\ref{C1aley2}). 

Before to continue it is worth to notice the relation between the variation of the mass parameter and the the horizon's radii defined by Eq.(\ref{RelacionHorizontes}). In this case these are given by
\begin{equation} \label{RelacionHorizontes1}
 dM= T_+ nr_+^{d-2n-1}\left (\frac{\partial m(r_+,M)}{\partial M} \right)^{-1} dr_+ 
   = - T_{++} nr_{++}^{d-2n-1}\left (\frac{\partial m(r_{++},M)}{\partial M} \right)^{-1} dr_{++} ,
\end{equation}
and therefore one can argue, as there is a single mass parameter, that as $M$ increases then $r_+$ must increase but $r_{++}$ must decrease, and viceverse. This will be very useful in the next sections to connect the thermodynamic evolution with the evolution of the horizons $r_+$ and $r_{++}$.

\section{Lovelock theory with $n$-fold degenerated ground state as a thermodynamic identity}

The asymptotically locally AdS vacuum solutions of the $n$-degenerated ground state Lovelock have been studied in references \cite{Crisostomo:2000bb,Aros:2000ij}. The regular solution for negative and positive cosmological constant were obtained in \cite{milko}, to our knowledge for first time. The corresponding equation of motion reduces to solve
\begin{equation}\label{eqmotionBHS}
 \frac{d}{dr}\left(r^{d-2n-1}\left[1 - f(r,M) - \frac{r^2}{l^2}\right]^{n}\right) = \rho(r) \Omega_{d-2} r^{d-2},
\end{equation}
whose solution is
\begin{equation} \label{solucionBHS}
    f(r,M) = 1 - \frac{r^2}{l^2}- \left( \frac{m(r,M)}{r^{d-2n-1}} \right)^{1/n}.
\end{equation}
Here $d-2n-1>0$. The mass function $m(r,M)$ is defined by the equation (\ref{massfunction}) as previously. If the mass function fulfills the criteria listed in reference \cite{milko} the solution (\ref{solucionBHS}) also can be a regular black hole. Moreover, the structure zeros of $f(r,M)$ in this case can be obtained by drawing the pair of functions
\begin{equation} \label{zero2}
\left( 1-\frac{r^{2}}{l^{2}}, \left(\frac{m(r,M)}{r^{d-2n-1}}\right)^{1/n} \right),
\end{equation}
whose intersection defines the zero. One can notice that this shares the generic behavior seen in Fig.(\ref{zerostructure}). Therefore, there are at most three horizons, $r_{-} < r_{+} < r_{++}$ for a range of mass parameter $M_{cri1}<M<M_{cri2}$, there are two different scenarios where only one horizon exists, and there are extreme solutions with two horizons for $M=M_{cri1}$ and $M=M_{cri2}$. 

\subsection{Local Thermodynamics}

To deduce the local thermodynamics one must consider a horizon defined by $r=a$. Therefore, from Eq.(\ref{solucionBHS}) one can obtain 
\begin{equation}
    m(a)=  a^{d-2n-1} \left( 1 - \frac{a^2}{l^2} \right)^n.
\end{equation}
It is worth to stress that $m(M,r)$, once evaluated on a horizon, becomes just a function of the horizon radius. Now, by evaluating the equation of motion (\ref{eqmotionBHS}) at $r=a$ and $\rho=-P$ and multiplying by $da$ yields
\begin{equation} \label{primeraleyBHS}
    P d \left( \frac{\Omega_{d-2}a^{d-1}}{d-1} \right) = \frac{f'}{4\pi} \left( 4 \pi n  a^{d-2n-1} \Big [ 1-\frac{a^2}{l^2} \Big ]^{n-1} \right)da  - d \left( m(a) \right).
\end{equation}
In this expression one can recognize the definition of variation of entropy 
\begin{equation}
    dS  =  4\pi n a^{d-2n-1} \left[ 1-\frac{a^2}{l^2} \right]^{n-1} da,
\end{equation}
obtained in \cite{Aros:2000ij}. 

The identification of Eq.(\ref{primeraleyBHS}) as a first law of thermodynamics is straightforward. As previously, at the black horizon, $a=r_{+}$, $f'|_{r=r_+}>0$, and thermodynamic pressure, volume, temperature and energy are defined analogously to Eqs.(\ref{presionBH},\ref{volumenBH},\ref{TemperaturaBH}) and (\ref{energiaBH}), respectively. Therefore, Eq.(\ref{primeraleyBHS}) takes the form $PdV=TdS-dU$. By the same token, at the cosmological horizon, $a=r_{++}$, is satisfied $f'|_{r=r_{++}}<0$. The thermodynamic pressure, volume, temperature and energy are defined in as in Eqs.(\ref{presionCH},\ref{volumenCH},\ref{TemperaturaCH}) and (\ref{energiaCH}), respectively,  Finally, Eq.(\ref{primeraleyBHS}) again can be rewritten as $P(-dV)=TdS-dU$. 

The heat capacities are given by 
\begin{eqnarray}
      C_+ &=&  r_{+}^{d-2n-1}\left(1 - \frac{r_+^2}{l^2}\right)^{n-1}\left( \frac{d-2n-1}{r_+} - \frac{(d-1)r_+}{l^{2}} \right) \left(\frac{\partial T_+}{\partial r_+} \right)^{-1} \label{C1aley3} \\
    C_{++} &=& r_{++}^{d-2n-1}\left(1 - \frac{r_{++}^2}{l^2}\right)^{n-1}\left(- \frac{d-2n-1}{r_{++}} + \frac{(d-1)r_{++}}{l^{2}} \right) \left(\frac{\partial T_{++}}{\partial r_{++}} \right)^{-1}, \label{C1aley4}
\end{eqnarray}
Here one can recognize the third factor as the previously known definition of the temperature of the vacuum solutions. This term is always positive. Therefore the sign of the heat capacity of the each horizon depends only on the sign of the factor  $\partial T/ \partial a$.

Again, from equation (\ref{HEATCAPACITY1}), and following the convention of reference \cite{AplicacionDS3}, the $K$-{\it heat capacity} is given by
\begin{equation}\label{nFoldHeat}
     K_{+}=  4 \pi n r_{+}^{d-2n-1} \left ( 1-\frac{r_{+}^2}{l^2}  \right )^{n-1}  T_+  \left. \left ( \frac{\partial m(r,M)}{\partial M} \right )^{-1}\right|_{r=r_{+}}  \left(\frac{\partial T_+}{\partial r_{+}} \right)^{-1}.
\end{equation}
and analogously for $r_{++}$. One can notice that the four first factors are positive for both $a=r_+$ and $a=r_{++}$. Therefore, the sign of the $K_+$ and $K_{++}$ depends only on the sign of  $\partial T/ \partial a$.

Before to continue it is worth to notice the relation between the variation of the mass parameter and the the horizon's radii defined by Eq.(\ref{RelacionHorizontes}). In this case these are given by 
\begin{eqnarray} \label{RelacionHorizontes22}
 \frac{dM}{dr_+}&=& T_+ nr_+^{d-2n-1} \left (1-\frac{r_+^2}{l^2} \right )^{n-1}\left (\frac{\partial m(r_+,M)}{\partial M} \right)^{-1} \nonumber \\
  \frac{dM}{dr_{++}} \ &=& - T_{++} nr_{++}^{d-2n-1} \left (1-\frac{r_{++}^2}{l^2} \right )^{n-1} \left (\frac{\partial m(r_{++},M)}{\partial M} \right)^{-1}.
\end{eqnarray}
As for the Pure Lovelock solution, in this case one can argue that as $M$ increases then $r_+$ must increase but $r_{++}$ must decrease, and viceverse.    

\section{A particular mass density} \label{ejemplo}
Although the analysis above was carried out without the use of an explicit mass density $\rho$, still one can consider to analyse one in particular to notice how the results above become manifest. Because of that, in this section it is analyzed a new $d$ dimensional generalization of the mass density proposed in \cite{Dymnikova1,Dymnikova2}:
\begin{equation} \label{DensidadDymnikova}
    \rho (r) = \frac{M}{V} \exp \left(  - \frac{r^{d-1}}{R^{d-1}} \right),
\end{equation}
where $M$ is the mass parameter. $R$ is a parameter of units of length $L$. $V$ is given by
\begin{equation}
    V= \frac{\Omega_{d-2}R^{d-1}}{d-1}.
\end{equation}

The direct integration of the mass density in Eq.(\ref{massfunction}) yields the mass function
\begin{equation} \label{MasaDymnikova}
    m(r)= M \left(  1 -  \exp \left(  - \frac{r^{d-1}}{R^{d-1}} \right) \right).
\end{equation}
The mass function Eq.(\ref{MasaDymnikova}) behaves as Eq.(\ref{Masacercaorigen}) for $r \ll R \ll l$, and therefore this defines proper  {\it Lovelock} regular black holes.

\subsection{Horizons}
\begin{figure}
\centering
\subfigure[$f(r)$ for $n=1,d=5$, $M_{cri1}=6$, $M_{cri2}=25$.]{\includegraphics[width=75mm]{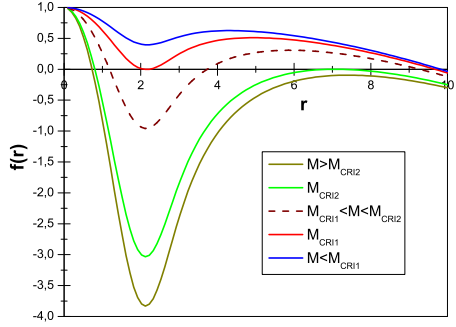}}
\caption{$f(r)$ of EH for and $R=2$ and $l=10$.}
\label{figurahorizontesEH}
\end{figure}

\begin{figure}
\centering
\subfigure[$f(r)$ for $n=2,d=6$, $M_{cri1}=2.6$, $M_{cri2}=5.4$.]{\includegraphics[width=75mm]{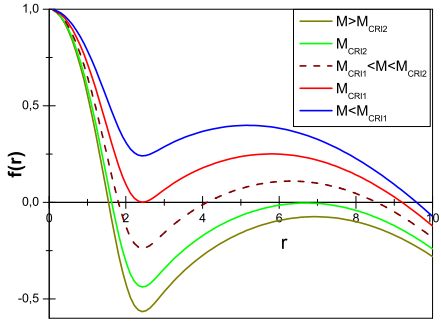}}
\subfigure[$f(r)$ for $n=3,d=8$, $M_{cri1}=2.5$, $M_{cri2}=6.2$.]{\includegraphics[width=75mm]{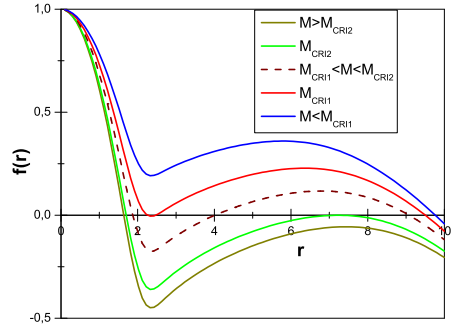}}
\caption{$f(r)$ of Pure Lovelock for and $R=2$ and $l=10$.}
\label{figurahorizontesPL}
\end{figure}

\begin{figure}
\centering
\subfigure[$f(r)$ for $n=2,d=6$, $M_{cri1}=2.3$, $M_{cri2}=2.88$.]{\includegraphics[width=75mm]{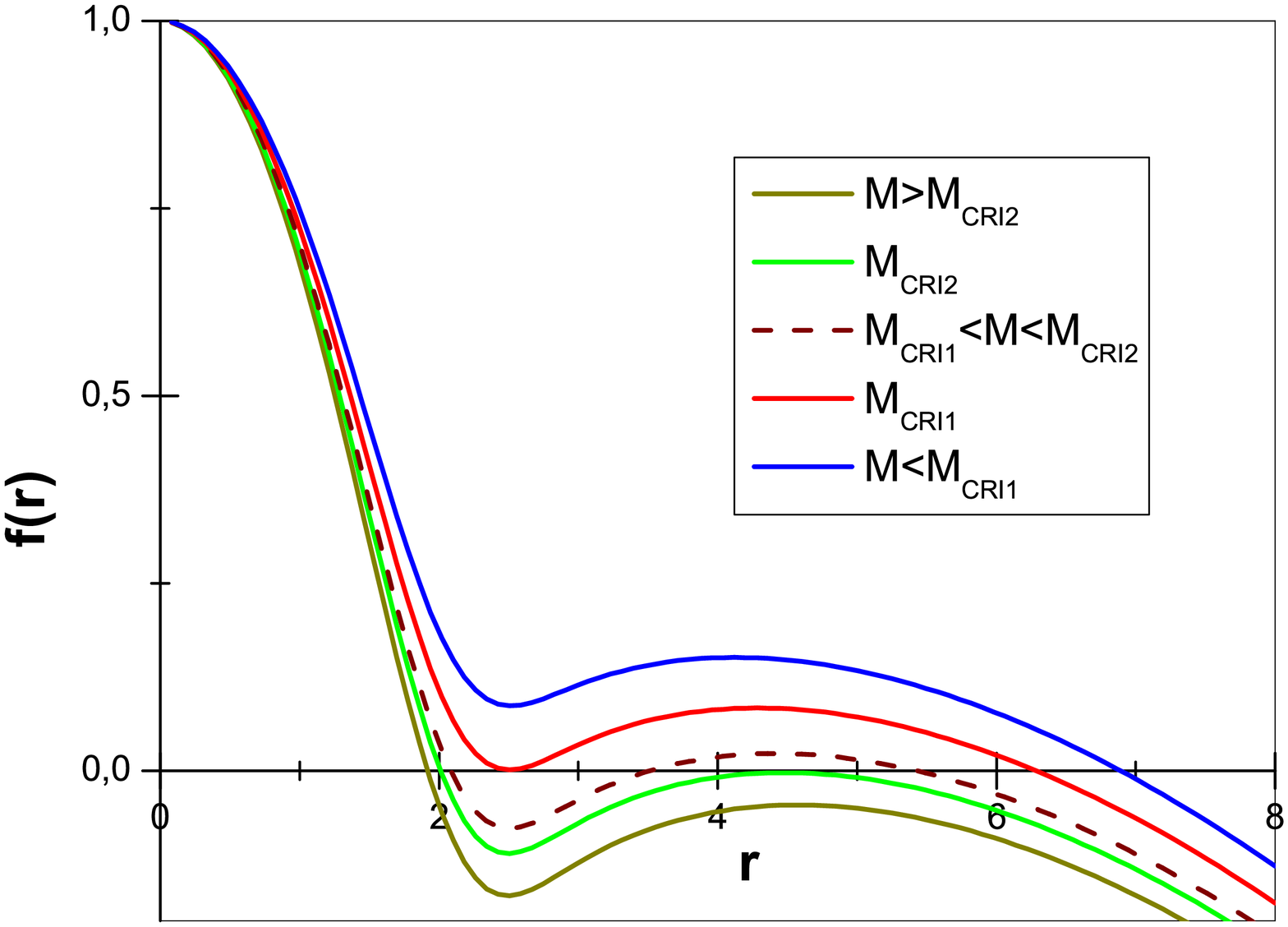}}
\subfigure[$f(r)$ for $n=3,d=8$, $M_{cri1}=2.05$, $M_{cri2}=2.4$.]{\includegraphics[width=75mm]{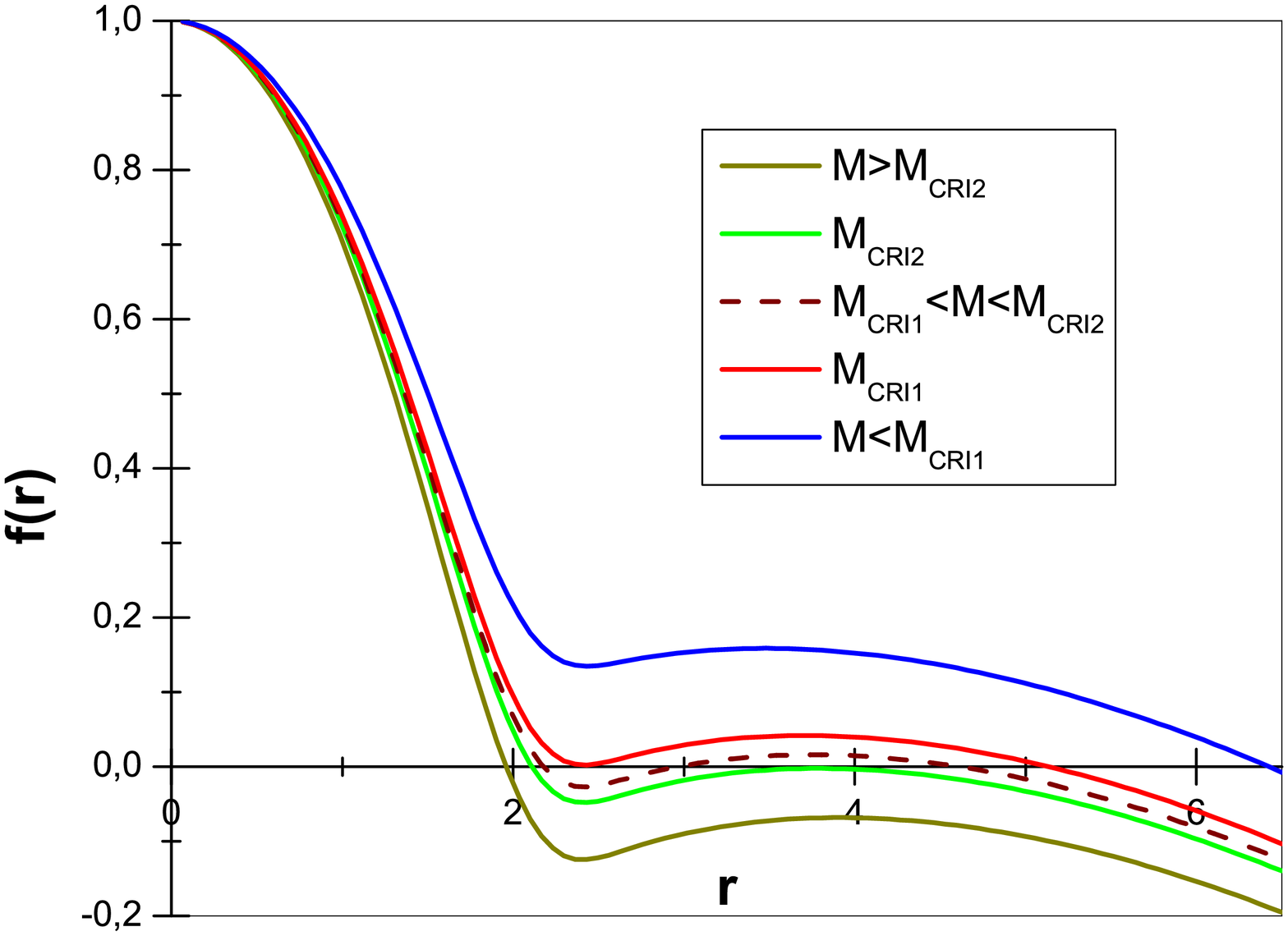}}
\caption{$f(r)$ of $n$-fold degenerated GS for and $R=2$ and $l=10$.}
\label{figurahorizontesBHS}
\end{figure}

Considering some toy values of $l$ and $R$, generic examples of the evolution of $f(r)$ are displayed in Fig. (\ref{figurahorizontesEH}) for the EH case, Fig.(\ref{figurahorizontesPL}) for the Pure Lovelock solutions, and Fig.(\ref{figurahorizontesBHS}) for the solutions with $n$-fold degenerated ground state. As mentioned above, one can notice that each of the solutions have a similar behavior as the one described in figure (\ref{zerostructure}).  Only within $M \in ]M_{cri1},M_{cri2}[$ there are three horizons and at the two critical values $M_{cri1}$ and $M_{cri2}$ the three generic horizons coalesce into just two horizons. Moreover, for $M<M_{cri1}$ and $M>M_{cri2}$ there is only one horizon. It is straightforward to check that this behavior is generic any other values of $d,l,n$.

\subsection{Evolution of $M$ parameter}


\begin{figure}
\centering
\subfigure[$f(r)=0$ for $n=1,d=5$]{\includegraphics[width=75mm]{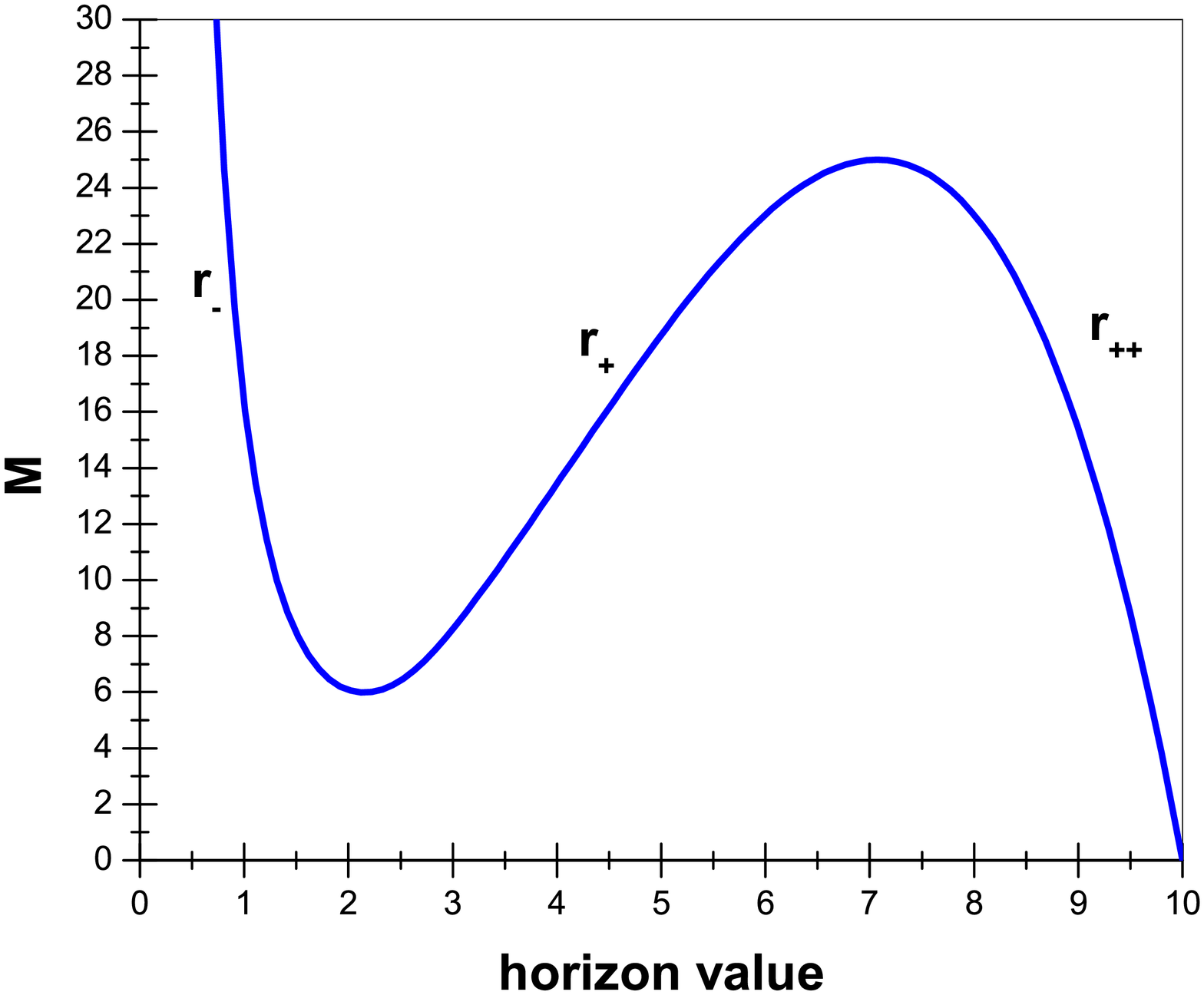}}
\caption{Behavior of parameter $M$ at the horizons in EH case.}
\label{figuraMEH}
\end{figure}

\begin{figure}
\centering
\subfigure[$f(r)=0$ for $n=2,d=6$]{\includegraphics[width=75mm]{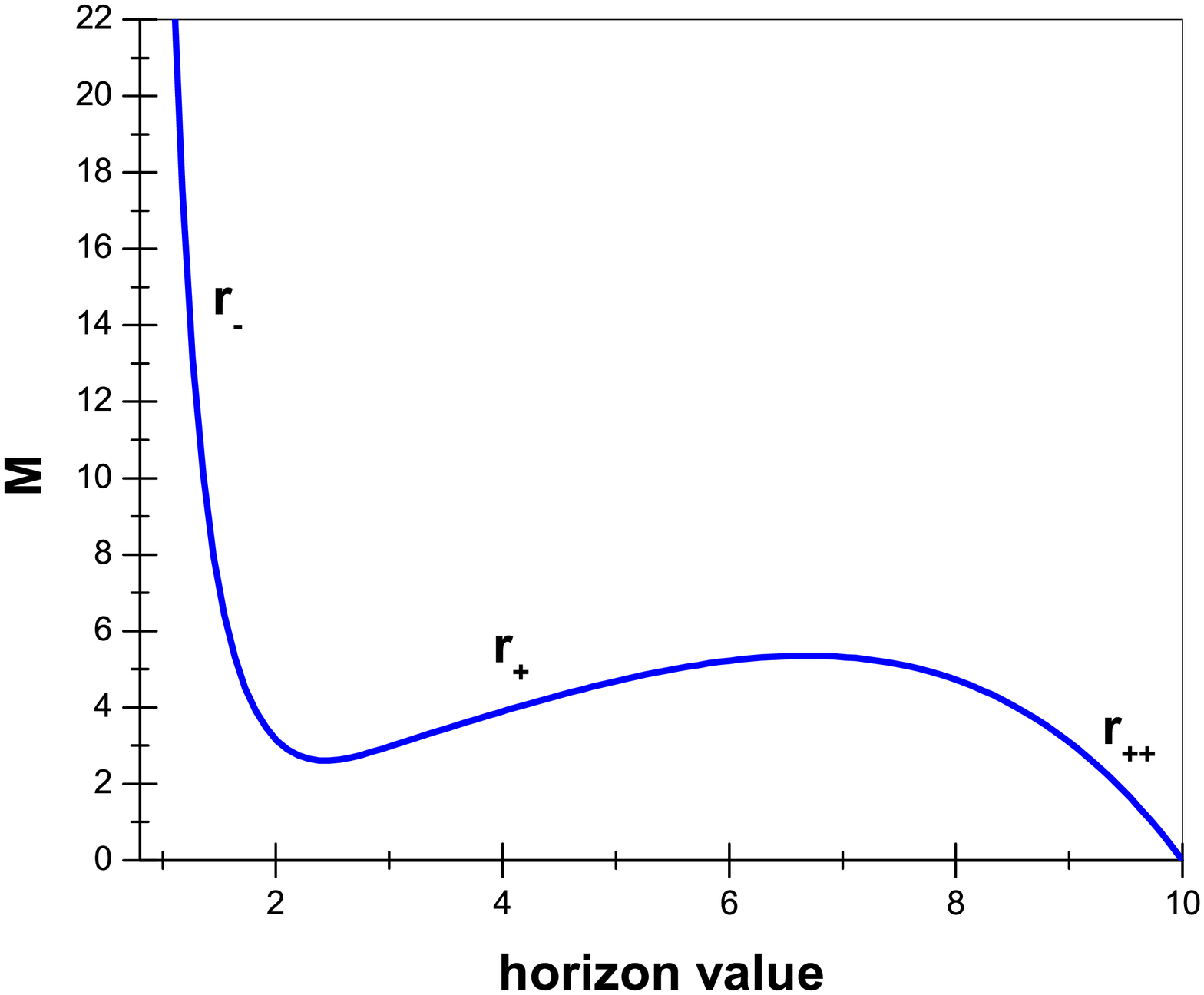}}
\subfigure[$f(r)=0$ for $n=3,d=8$.]{\includegraphics[width=75mm]{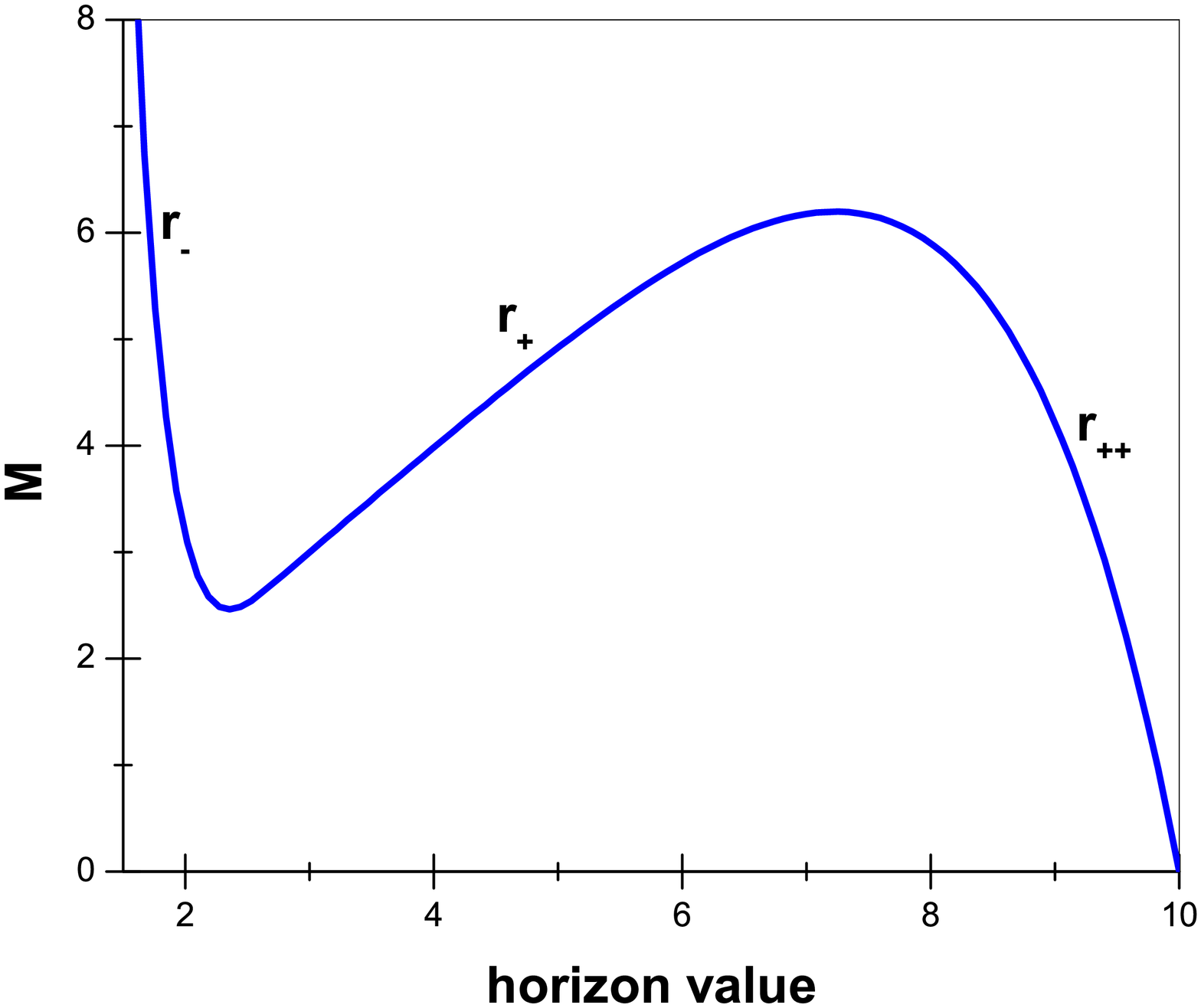}}
\caption{Behavior of parameter $M$ at the horizons in PL case.}
\label{figuraMPL}
\end{figure}

\begin{figure}
\centering
\subfigure[$f(r)=0$ for $n=2,d=6$.]{\includegraphics[width=75mm]{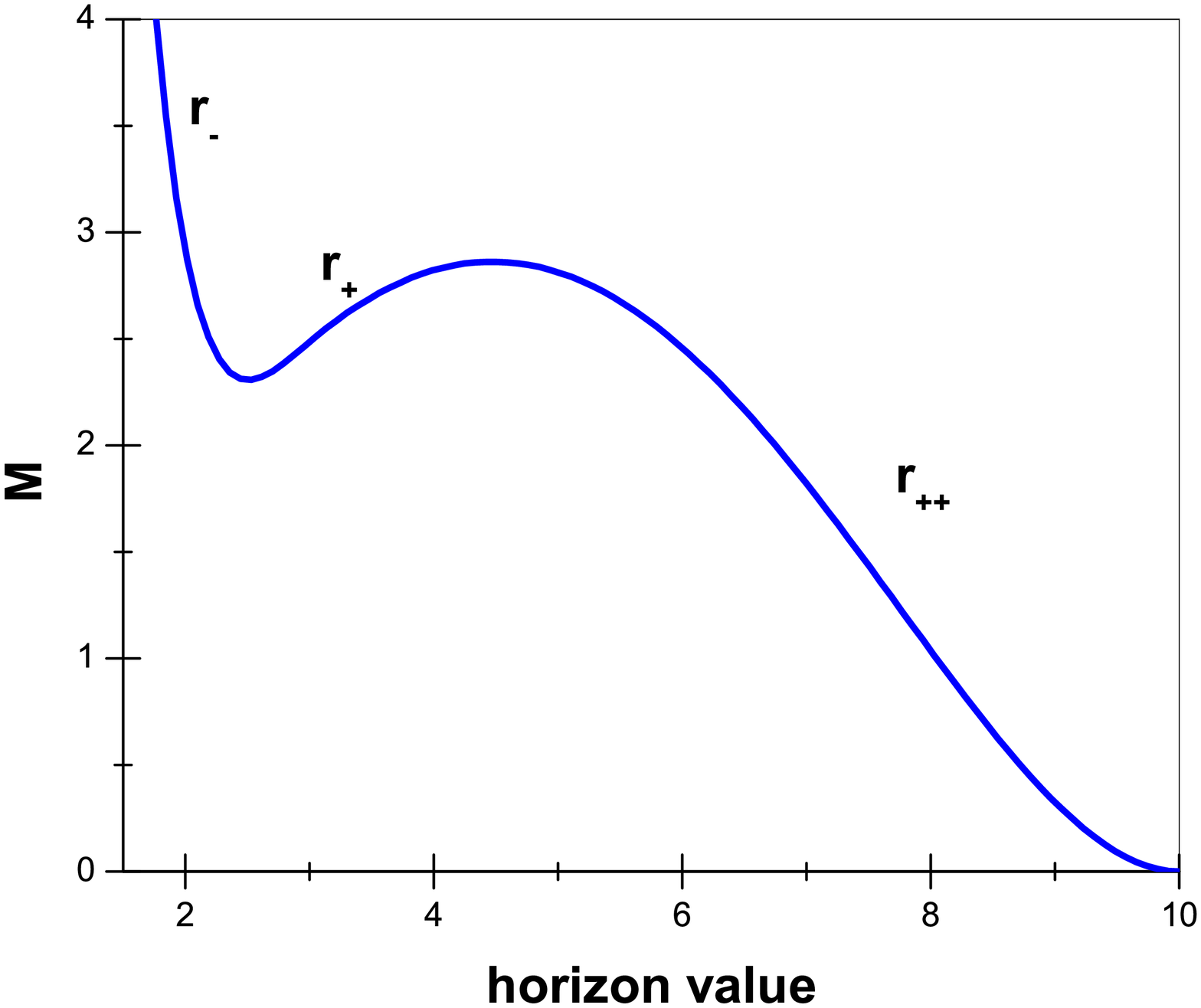}}
\subfigure[$f(r)=0$ for $n=3,d=8$.]{\includegraphics[width=75mm]{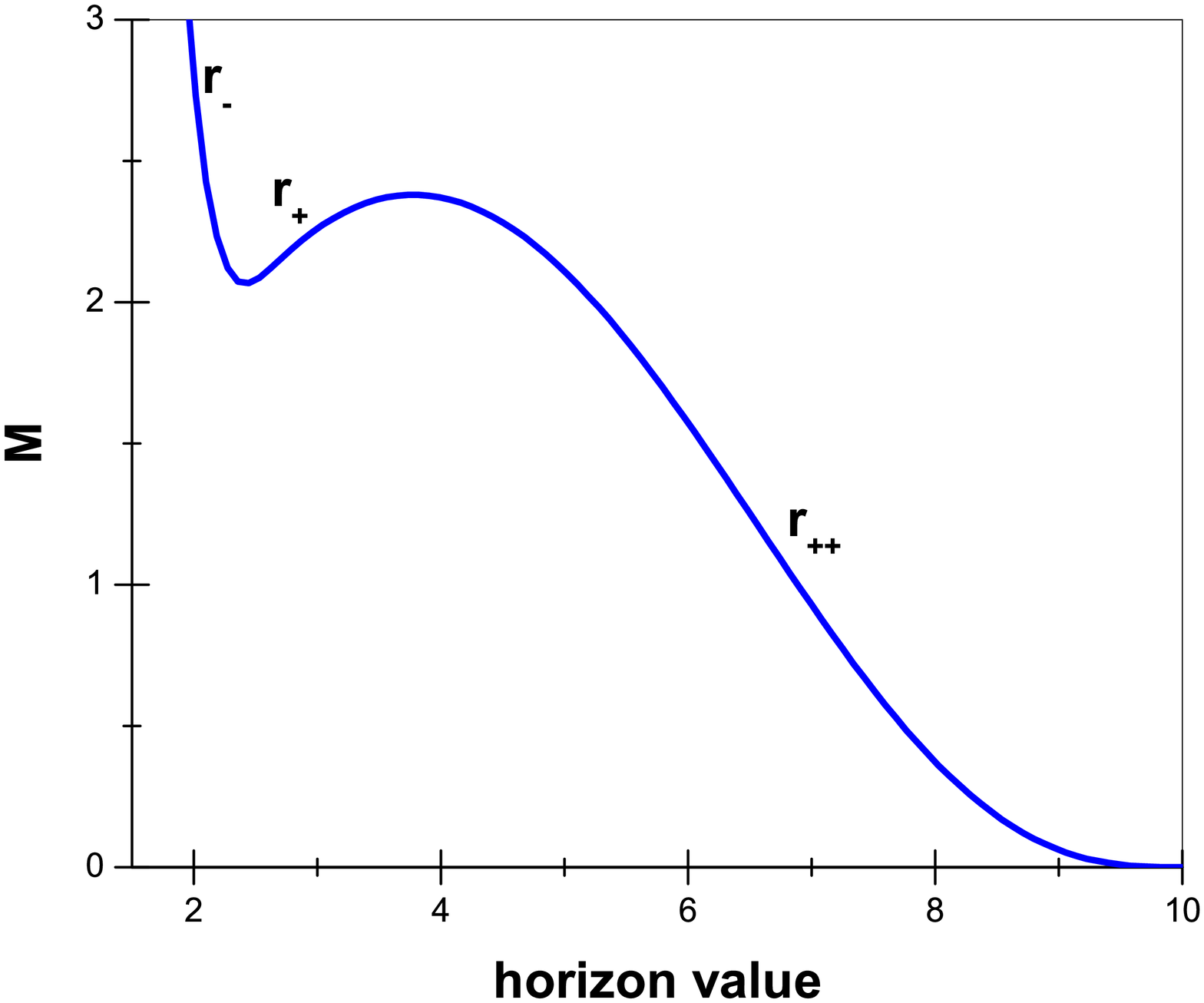}}
\caption{Behavior of $M$ at the horizons for $n$-fold degenerated GS solution.}
\label{figuraMBHS}
\end{figure}

The generic evolution of the $M$ parameter as a function of the horizon radii is showed in Fig.(\ref{figuraMEH}) for EH solution, Fig.(\ref{figuraMPL}) for Pure Lovelock solution and Fig.(\ref{figuraMBHS}) for the solution of the $n$-fold degenerated GS Lovelock. As previously, these are particular examples, but the same behavor could have be obtained for any $d$. One can notice that
\begin{equation}\label{MassesAsFunction}
  \frac{d M}{dr_{-}} \leq 0   \textrm{ , } \frac{d M }{d r_{+}} \geq 0  \textrm{ and } \frac{d M}{d r_{++}} \leq 0 
\end{equation}
which implies that as the mass decreases $r_+$ must decrease but $r_{++}$ must increase. 

\subsection{Energy at the horizons}
\begin{figure}
\centering
\subfigure[Mass function for $n=1,d=5$]{\includegraphics[width=75mm]{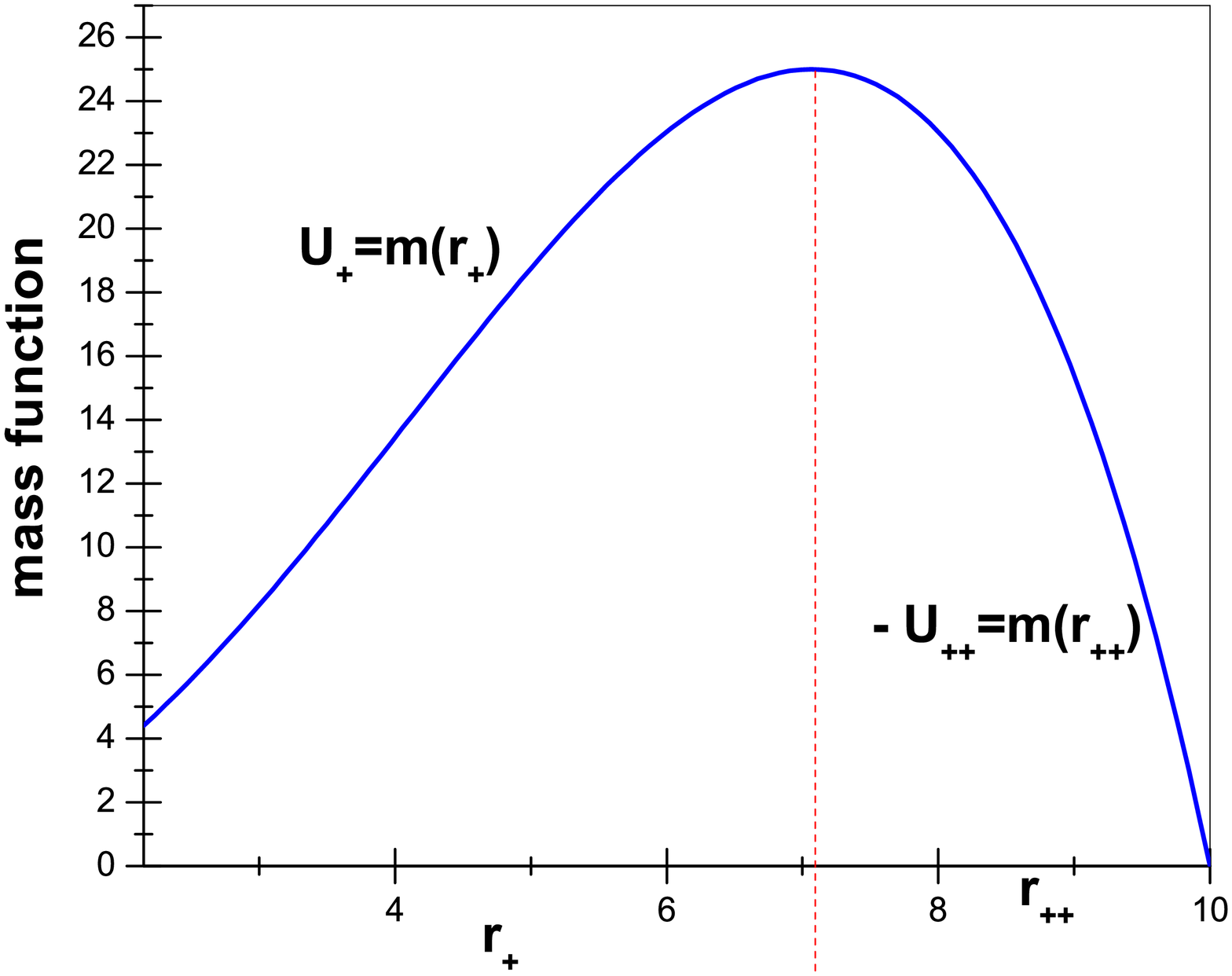}}
\caption{Behavior of mass function at the horizons in EH case.}
\label{figuraMASSEH}
\end{figure}

\begin{figure}
\centering
\subfigure[Mass function for $n=2,d=6$]{\includegraphics[width=75mm]{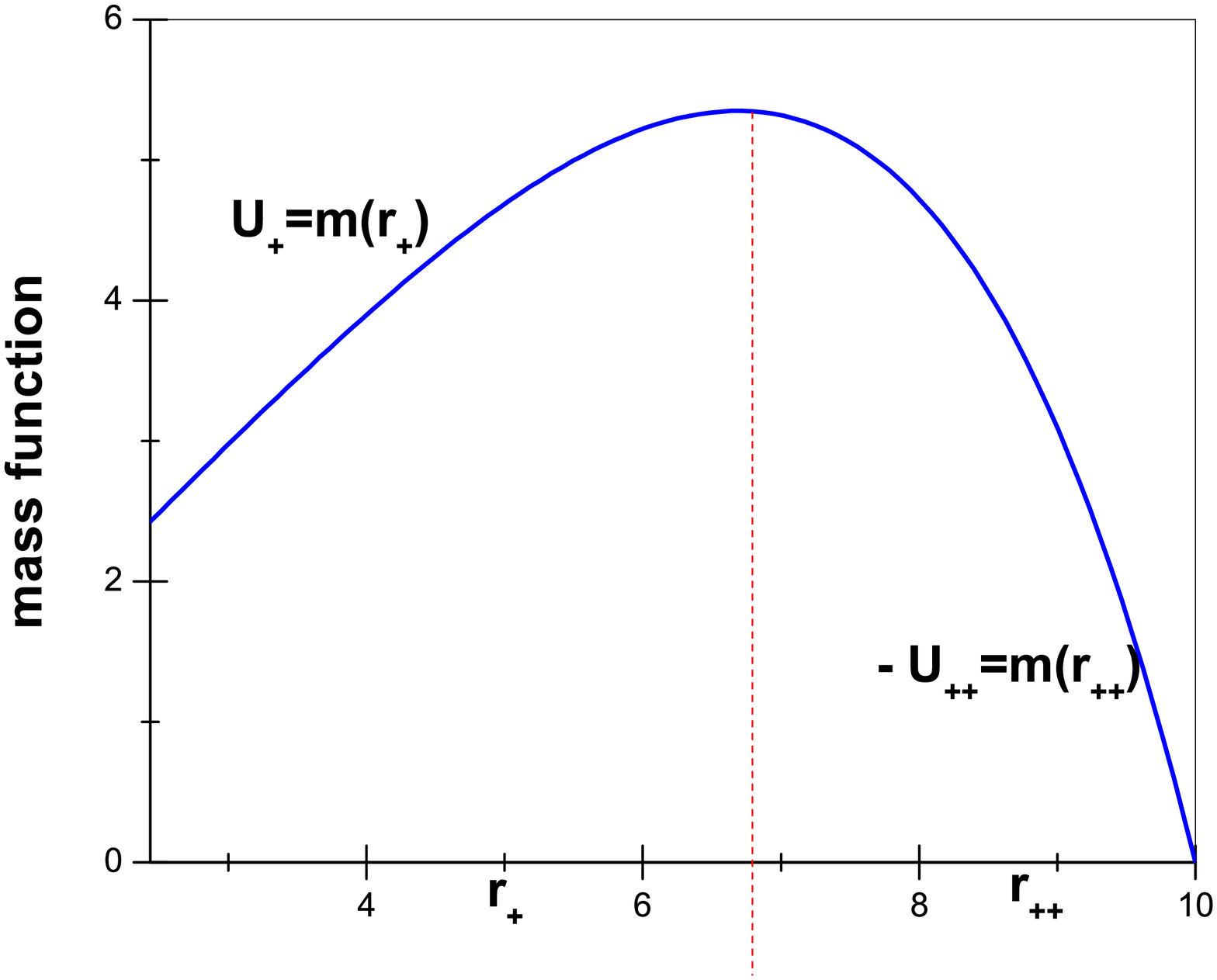}}
\subfigure[Mass function for $n=3,d=8$.]{\includegraphics[width=75mm]{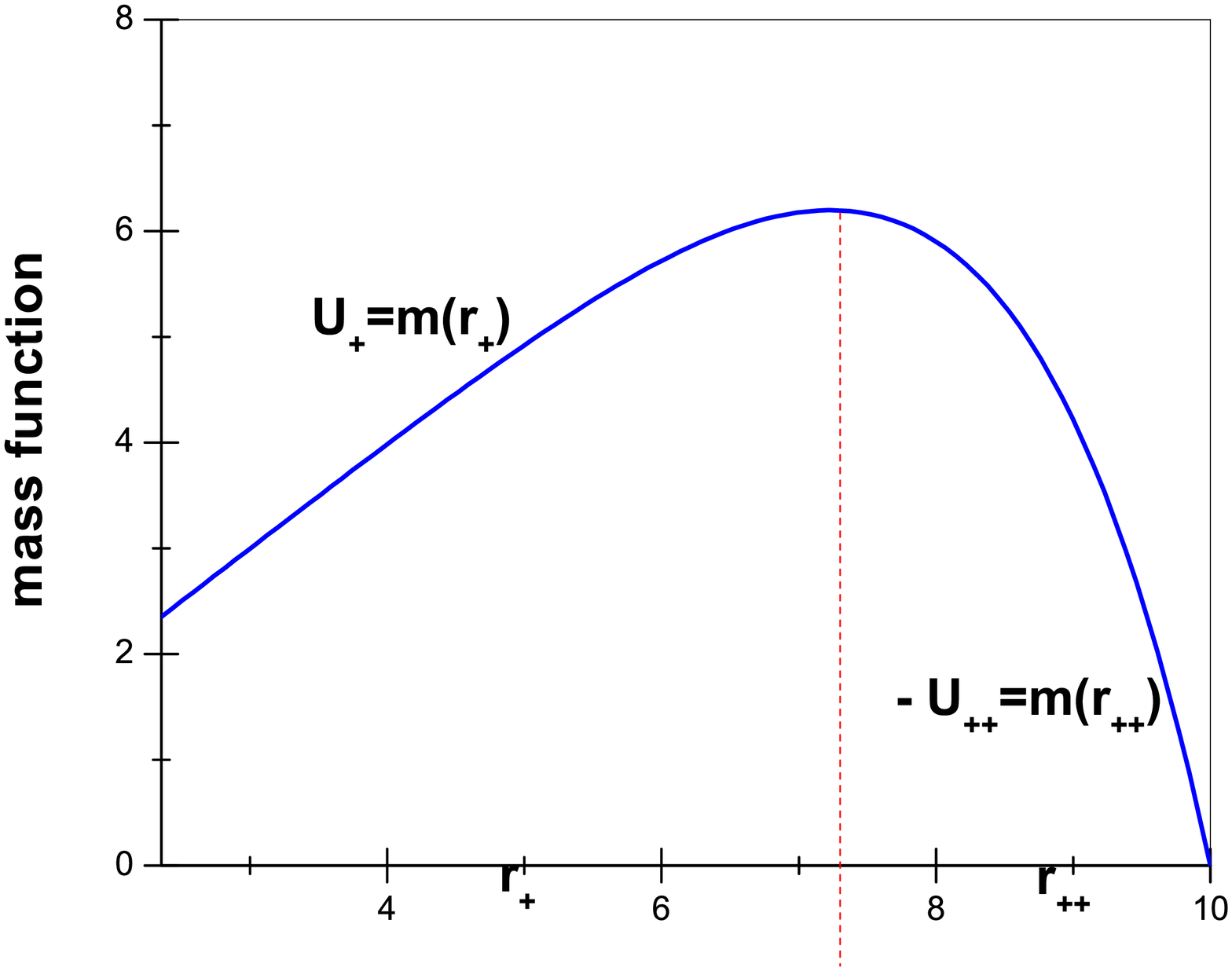}}
\caption{Behavior of mass function at the horizons in PL case.}
\label{figuraMASSPL}
\end{figure}

\begin{figure}
\centering
\subfigure[Mass function for $n=2,d=6$.]{\includegraphics[width=75mm]{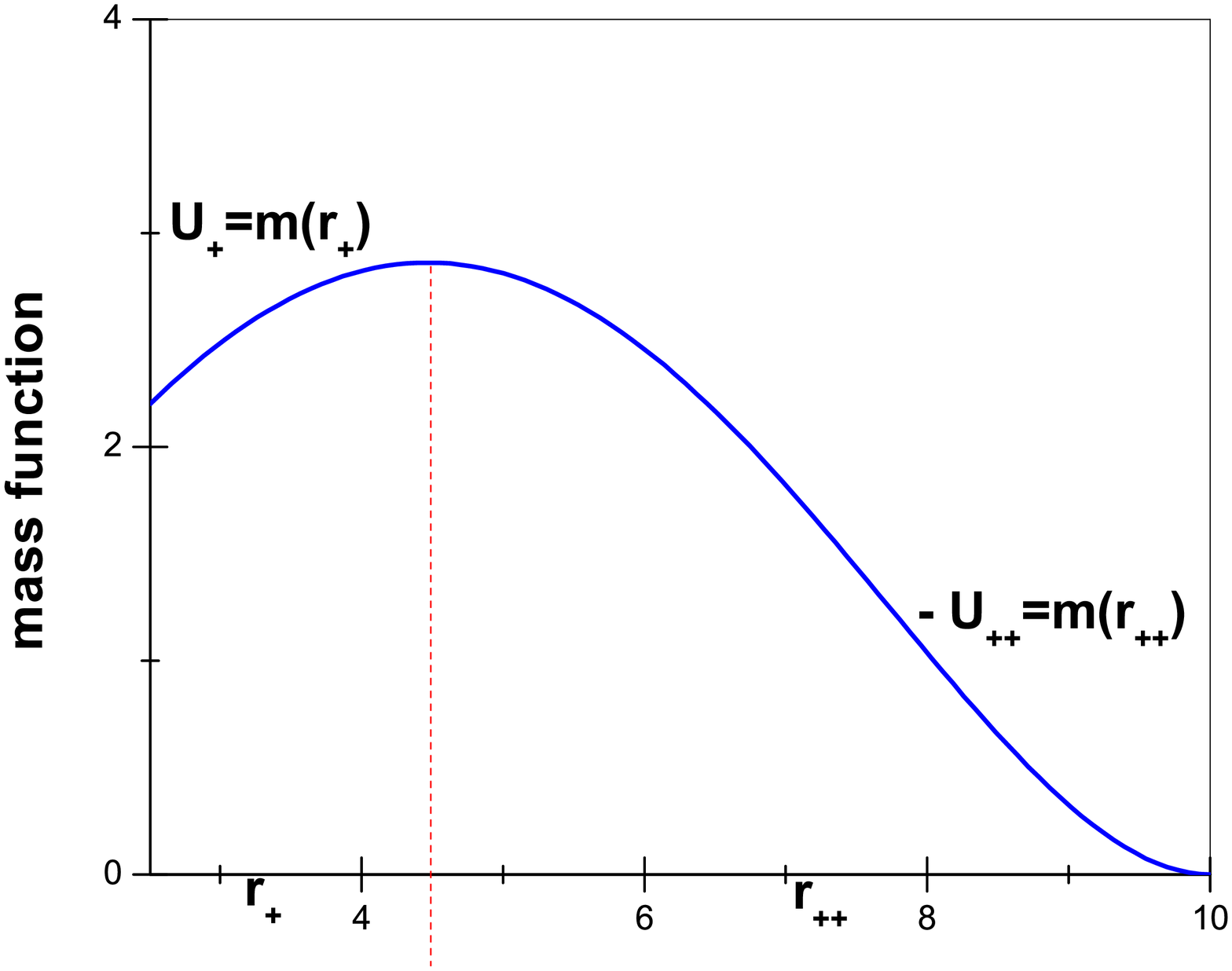}}
\subfigure[Mass function for $n=3,d=8$.]{\includegraphics[width=75mm]{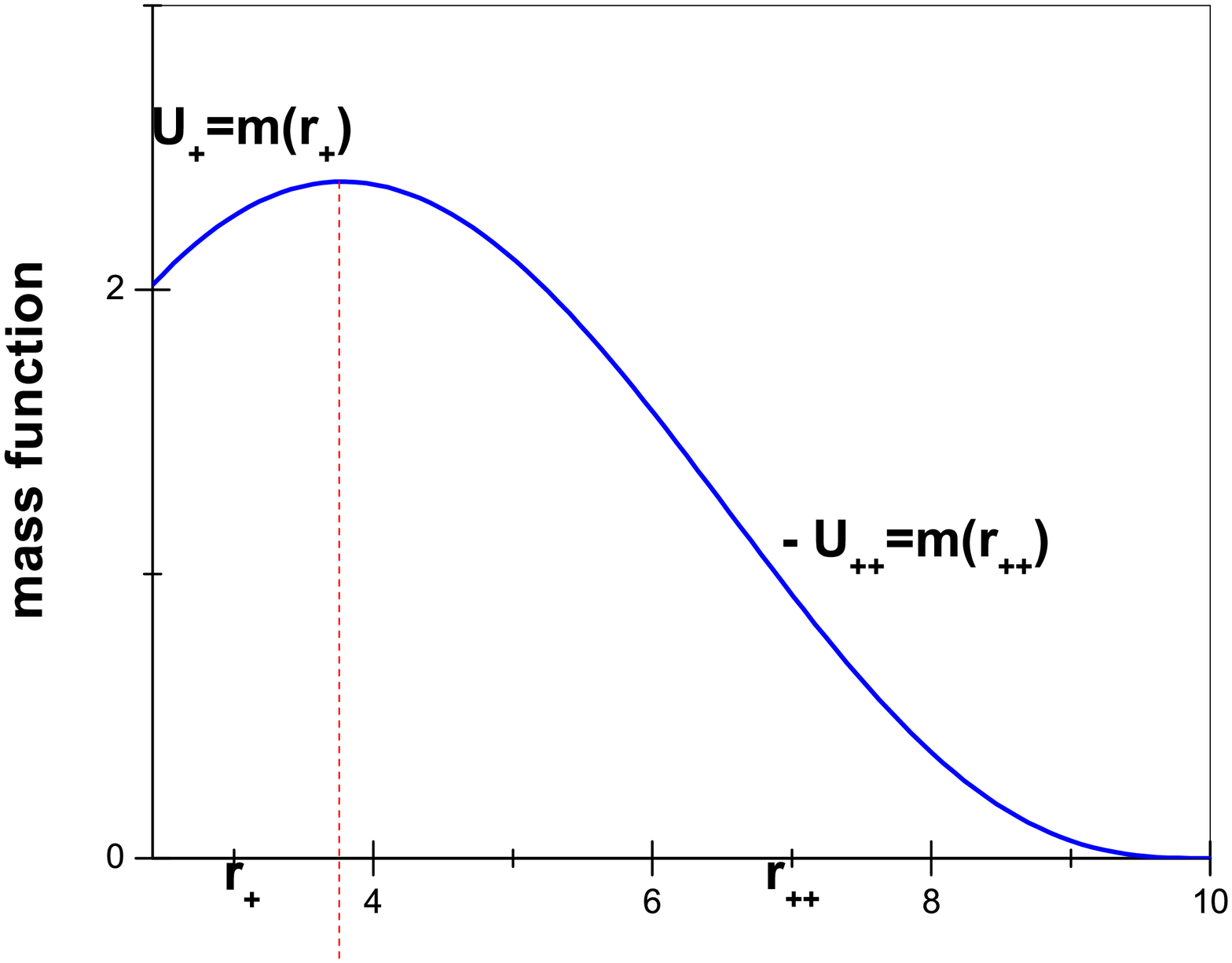}}
\caption{Behavior mass function at the horizons for $n$-fold degenerated GS solution.}
\label{figuraMASSBHS}
\end{figure}

The next analysis to be done comprehends the change of the local definition of energy mentioned in Eqs. (\ref{energiaBH}, \ref{energiaCH}).  In Figs.(\ref{figuraMASSEH}, \ref{figuraMASSPL},  \ref{figuraMASSBHS}) are depicted, for EH, Pure Lovelock and $n$-fold degenerated GS Lovelock solutions respectively, the change of energy as a function of the horizon radii. It is direct to notice that  

\begin{equation}\label{Massr+r++}
\frac{d U(r_+) }{d r_{+}} > 0  \textrm{ and }\frac{d U(r_{++}) }{d r_{++}} >0
\end{equation}

Although the exact form of the figures above depends on the precise model considered, it is straightforward to show that this behaviour is generic for any other $\rho(r)$, as along as the  conditions discussed in \cite{milko} are satisfied. One can notice that the curve has a local maximum. This corresponds to $r_+ = r_{++}$ and therefore when the geometry has a external zero temperature horizon. In this case $C_+ = C_{++}$ vanish, which is in agreement with the usual notion of a zero temperature object.    

\subsection{Temperature} \label{secciontemperatura}

\begin{figure}
\centering
\subfigure[Temperature for $n=1,d=5$]{\includegraphics[width=75mm]{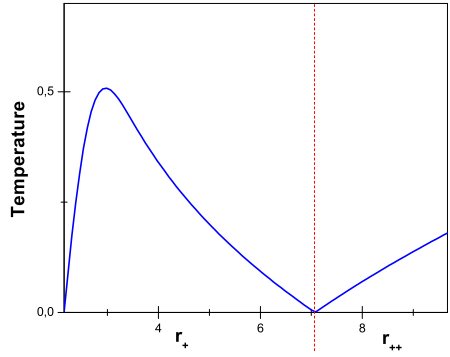}}
\caption{Behavior of temperature at the horizons in EH case.}
\label{figuraTEH}
\end{figure}

\begin{figure}
\centering
\subfigure[Temperature for $n=2,d=6$]{\includegraphics[width=75mm]{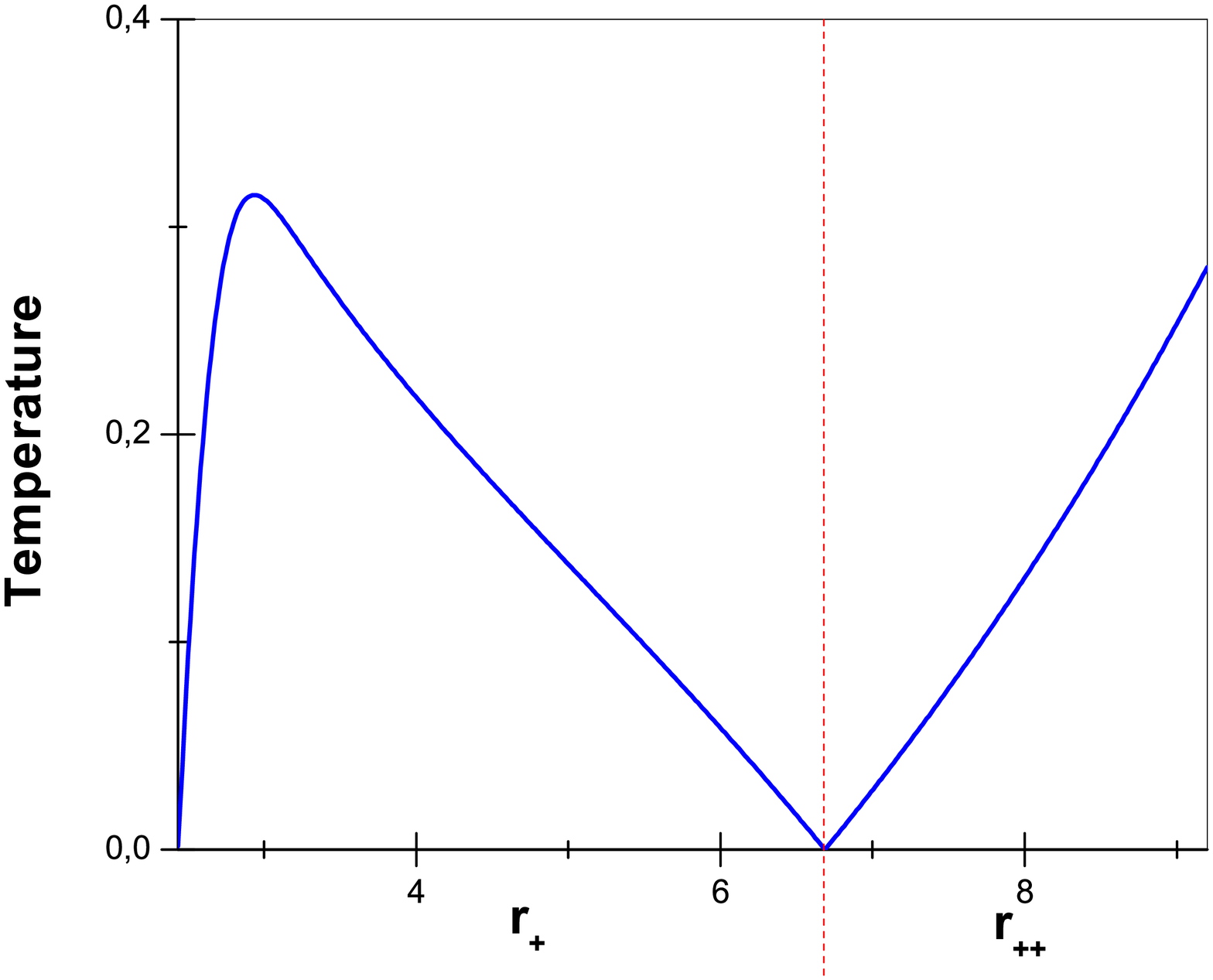}}
\subfigure[Temperature for $n=3,d=8$.]{\includegraphics[width=75mm]{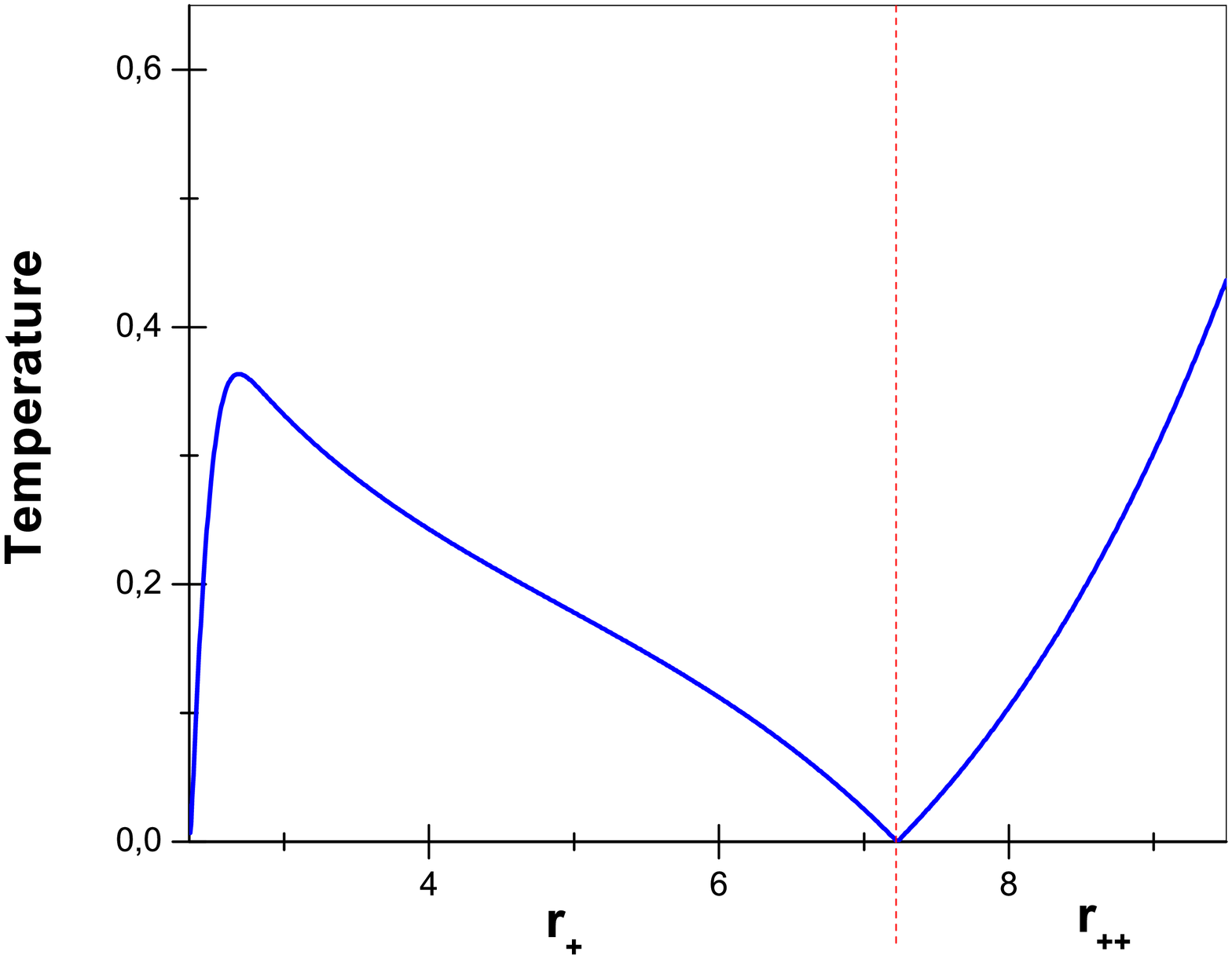}}
\caption{Behavior of temperature at the horizons in PL case.}
\label{figuraTPL}
\end{figure}

\begin{figure}
\centering
\subfigure[Temperature for $n=2,d=6$.]{\includegraphics[width=75mm]{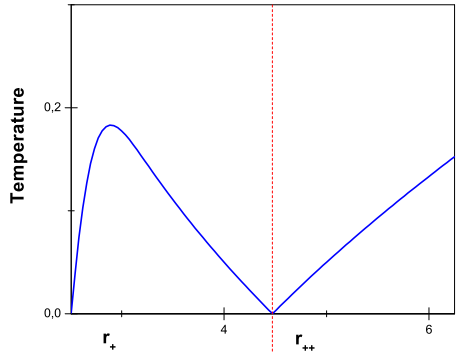}}
\subfigure[Temperature for $n=3,d=8$.]{\includegraphics[width=75mm]{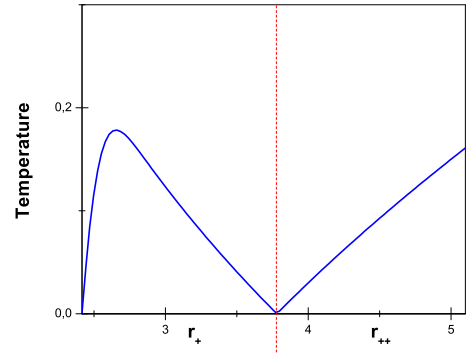}}
\caption{Behavior of temperature at the horizons for $n$-fold degenerated GS solution.}
\label{figuraTBHS}
\end{figure}

Figures (\ref{figuraTEH},  \ref{figuraTPL}) and (\ref{figuraTBHS}) display examples of the behavior of the temperature for EH, Pure Lovelock and $n$-fold degenerated GS solutions, respectively. In general one can notice the existence of a particular value $r_+ = r_*$ such that
\begin{equation}
    \frac{dT_+}{dr_+} = \left\{ \begin{array}{cc}
        > 0  &  r_+<r^* \\
        =0   &  r_+ = r^*\\
        < 0  &  r_+ > r^*
    \end{array} \right. 
\end{equation}
Therefore, temperature of black hole horizon reaches a local maximum at $r_+=r^*$. As $\frac{dT}{dr^*}$ vanishes at $r^*$ then the heat capacity $C_+=\frac{dU_+}{dr_+}/ \left(\frac{dT_+}{dr_+} \right)$ diverges in this point. See below. 

It must be noticed that, at cosmological horizon, the temperature is always an increasing function of $r_{++}$, therefore:
\begin{align}
& \frac{d T(r_{++}) }{d r_{++}} >0.\label{Tr++}
\end{align}

\subsubsection*{{\bf Black hole horizon temperature and cosmological horizon temperature:}}

\begin{figure}
\centering
\subfigure[$T_+$ v/s $T_{++}$ for $n=3,d=8$]{\includegraphics[width=75mm]{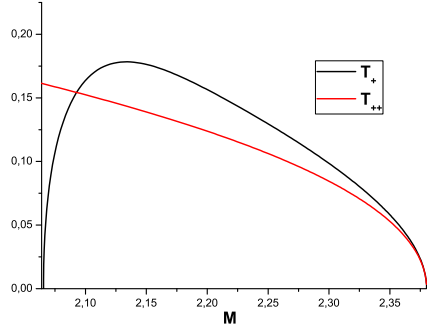}}
\caption{Behavior of $T_+$ v/s $T_{++}$ for $n$-fold degenerated GS solution.}
\label{DeltaTBHS}
\end{figure}

In order to simplify the discussion it is  convenient to define $M=M^*$ as the mass parameter associated with black hole horizon $r_+=r^*$ mentioned above. In figure (\ref{DeltaTBHS}), which corresponds to  $n$-fold degenerated GS solution with $n=3$ and $d=8$, that $T_+>T_{++}$ if $M>M^*$. However, there is one value $M=M_{eq}$, with  $M_{eq} \in ] M_{cri1}, M^*[$, where both temperatures reach the same value, $T_{+}(M_{eq}) = T_{++}(M_{eq}) \neq 0$. It is direct to check that this behavior is generic for the $n$-fold degenerated GS solutions for any other values of $d$ and $n$, including $n=1$ (Einstein Hilbert case).  

\begin{figure}
\centering
\subfigure[$T_+$ v/s $T_{++}$ for $n=2,d=6$]{\includegraphics[width=75mm]{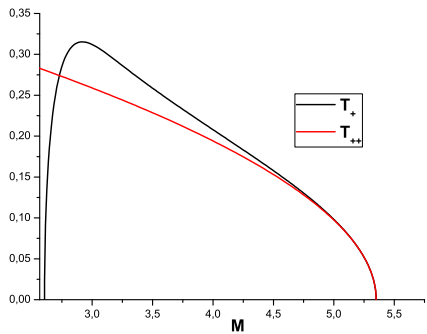}}
\subfigure[$T_+$ v/s $T_{++}$ for $n=3,d=8$]{\includegraphics[width=75mm]{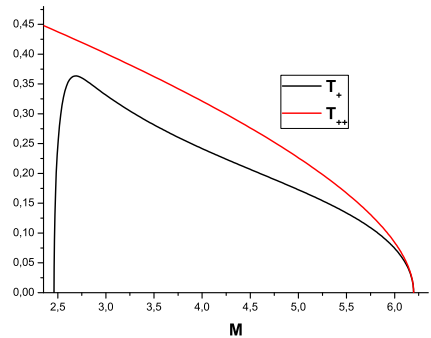}}
\subfigure[$T_+$ v/s $T_{++}$ for $n=3,d=9$]{\includegraphics[width=75mm]{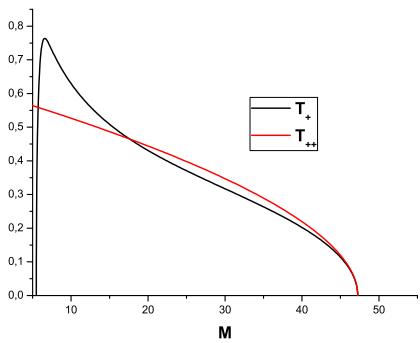}}
\caption{Behavior of $T_+$ v/s $T_{++}$ for Pure Lovelock.}
\label{DeltaTPL}
\end{figure}

The situation is different for the Pure Lovelock solutions. There are three scenarios, depending on the values of $n$, $d$ and $l$, 
\begin{itemize}
    \item The first case is one similar behavior that the Einstein Hilbert and $n$-fold degenerated solutions, where the $T_{+} > T_{++}$ for $M \in ]M_{eq},M_{cri2}[$ with $M_{cri1}<M_{eq}< M^*$. 
    \item The second case corresponds to $T_{+} < T_{++}$ for all value of $M \in [M_{cri1},M_{cri2}[$.
    \item The third case is when $T_{+} > T_{++}$ valid only for a range $[M_{eq},M_{CR}]$, with $M_{cri1}<M_{eq}<M^*$ and $M^*<M_{CR}<M_{cri2}$. $T_{+} < T_{++}$ for $M \in ]M_{CR}, M_{cri2}[$. 
   \end{itemize} 
Some numerical examples of the behavior of the Pure Lovelock case in figure (\ref{DeltaTPL}).

\subsection{Heat Capacity}
\begin{figure}
\centering
\subfigure[$C$ at BH horizon for $n=1,d=5$]{\includegraphics[width=75mm]{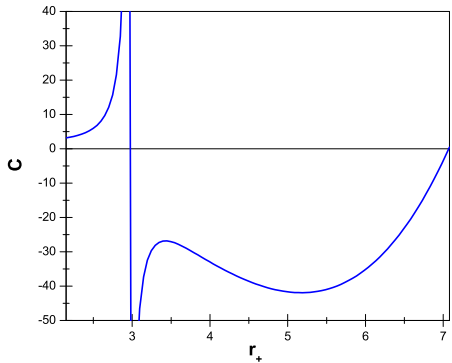}}
\caption{Behavior of heat capacity at BH horizon in EH case.}
\label{figuraCEH}
\end{figure}

\begin{figure}
\centering
\subfigure[$C$ at BH horizon for $n=2,d=6$]{\includegraphics[width=75mm]{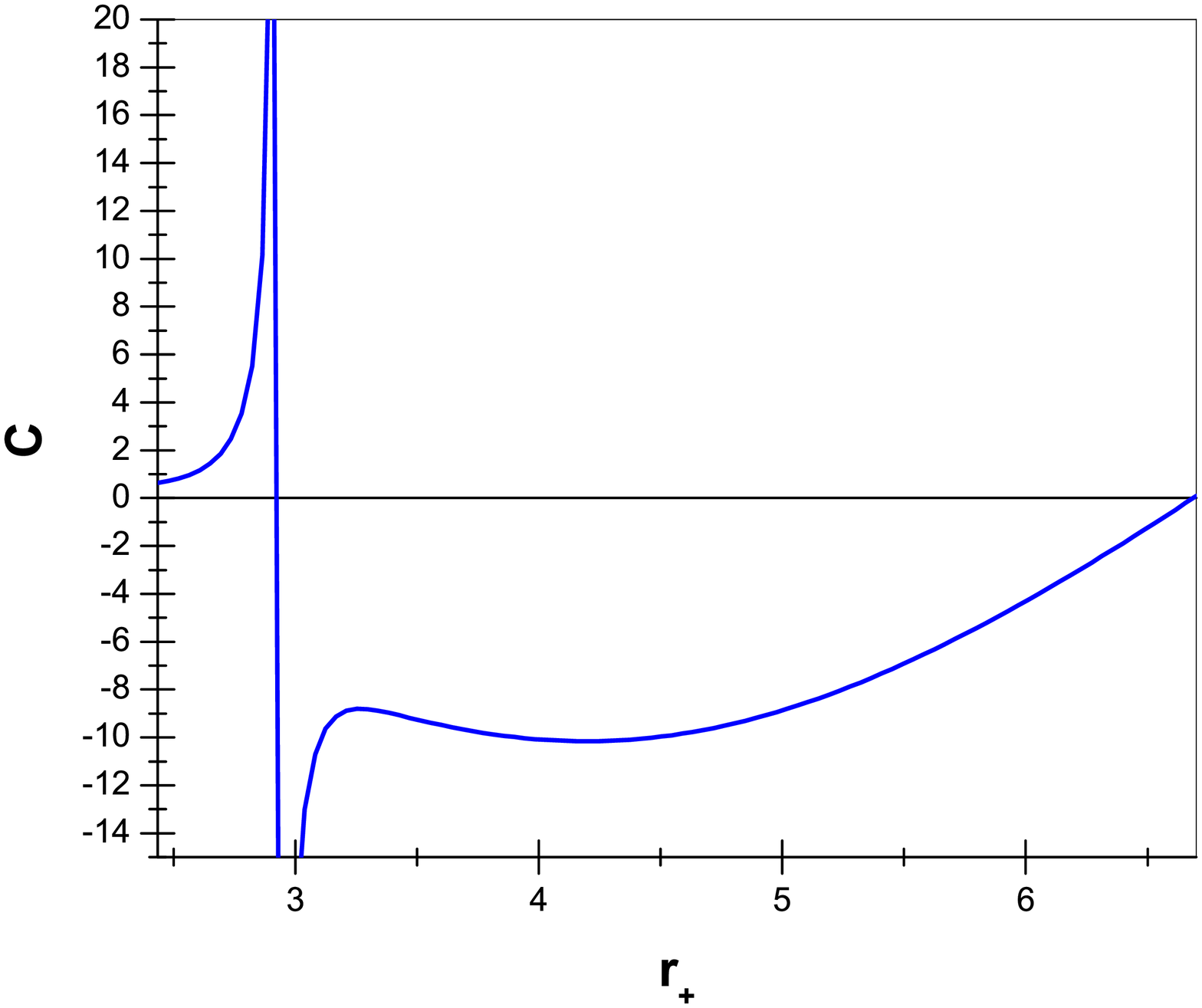}}
\subfigure[$C$ at BH horizon for $n=3,d=8$.]{\includegraphics[width=75mm]{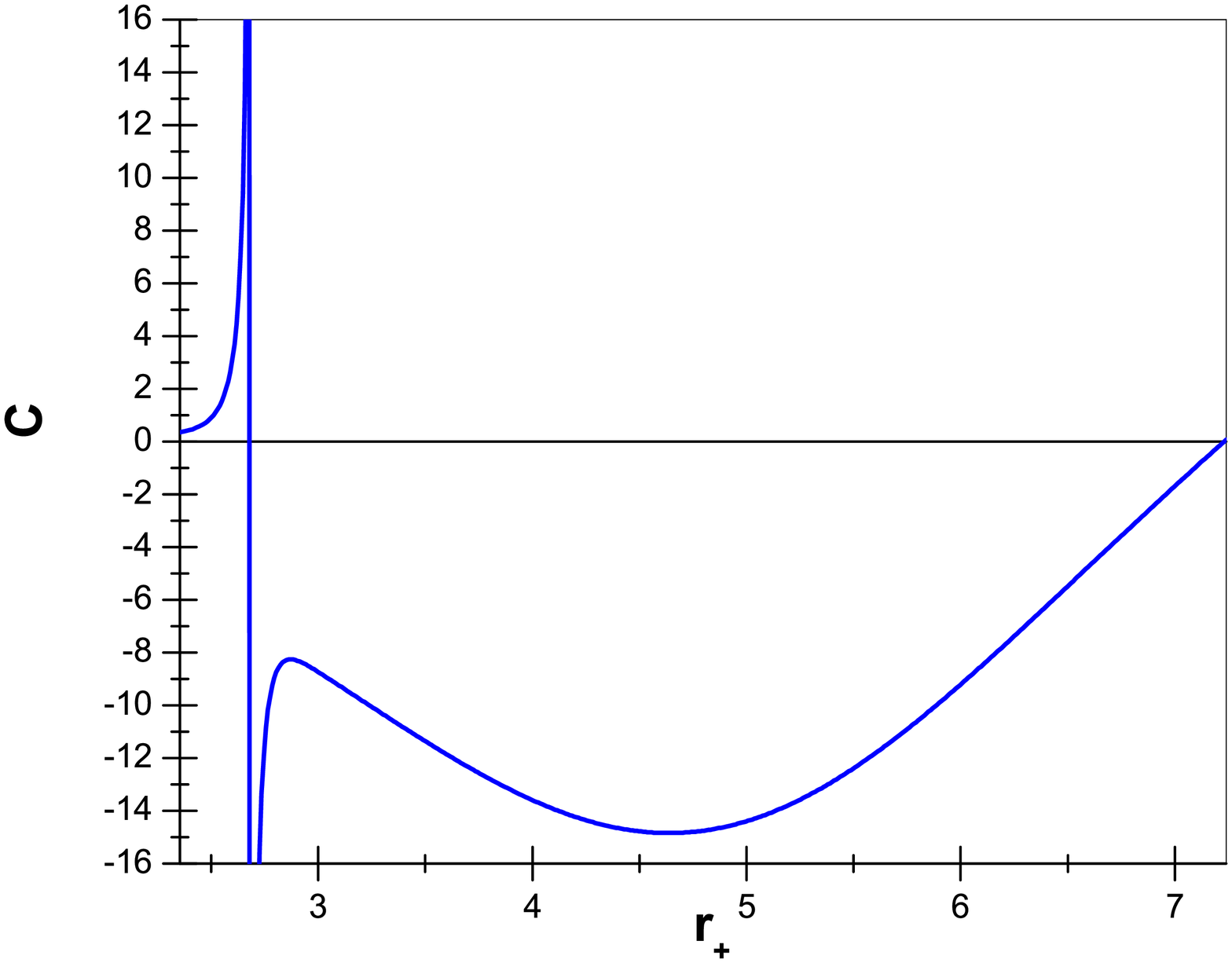}}
\caption{Behavior of heat capacity at BH horizon in PL case.}
\label{figuraCPL}
\end{figure}

\begin{figure}
\centering
\subfigure[$C$ at BH horizon for $n=2,d=6$.]{\includegraphics[width=75mm]{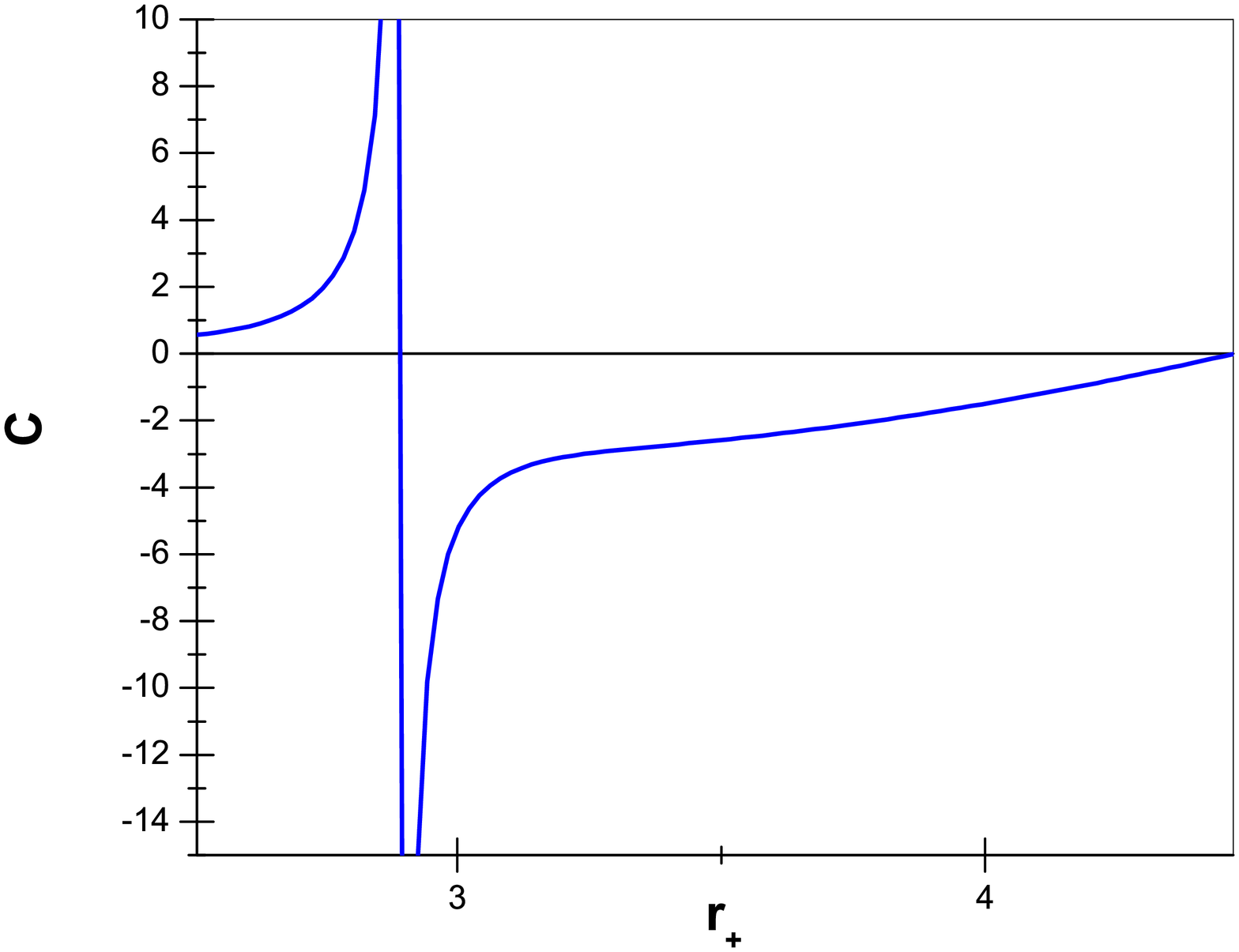}}
\subfigure[$C$ at BH horizon for $n=3,d=8$.]{\includegraphics[width=75mm]{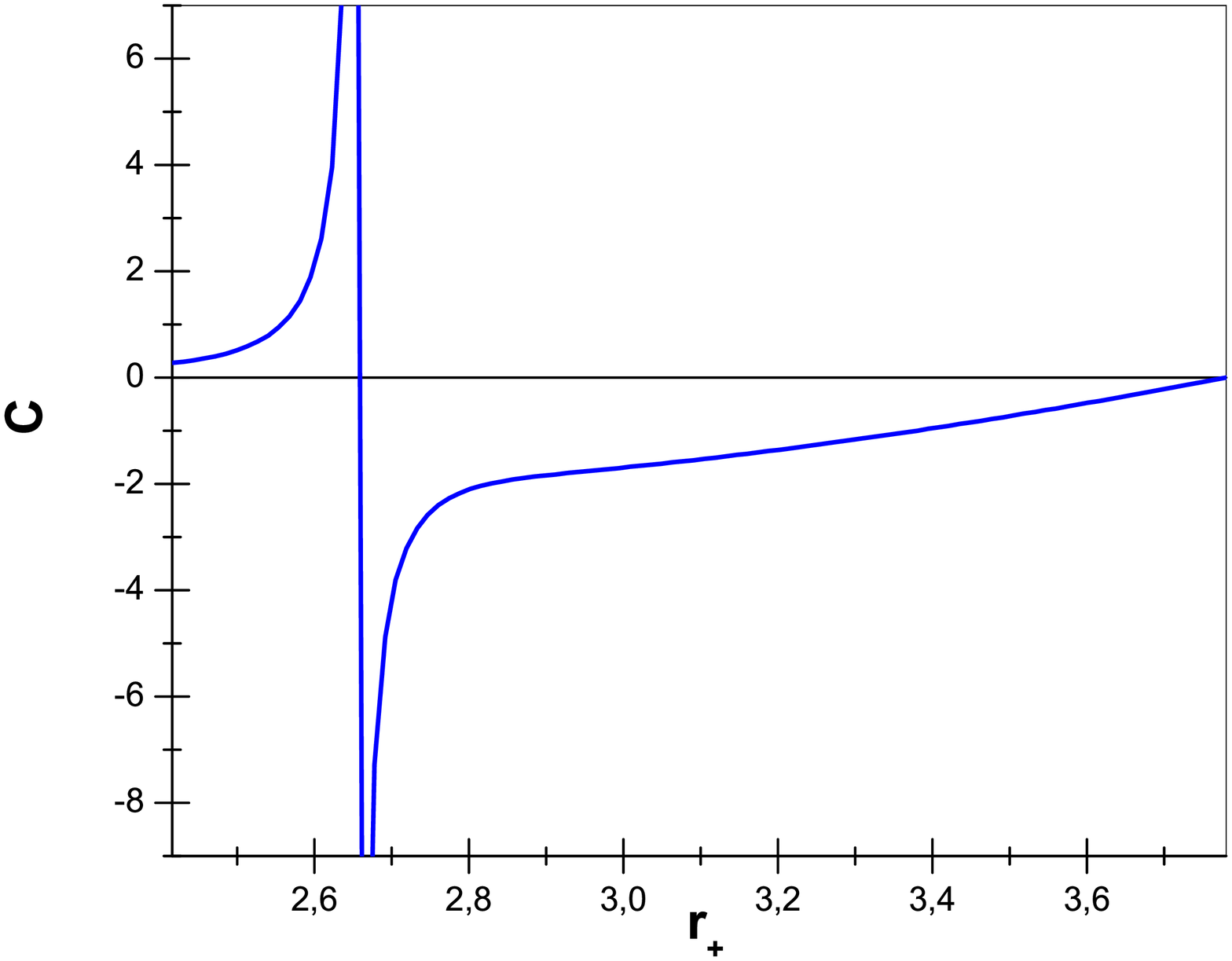}}
\caption{Behavior of heat capacity at BH horizon for $n$-fold degenerated GS solution.}
\label{figuraCBHS}
\end{figure}

First of all, it is worth stressing that at the cosmological horizon, by using equations (\ref{Massr+r++}) and (\ref{Tr++}), the {\it heat capacity} is always positive, i.e.,
\begin{equation}
    C_{++}=\frac{dU_{++}}{dr_{++}}/ \left(\frac{dT}{dr_{++}} \right) >0,
\end{equation}
therefore during the interaction the cosmological horizon would always aim to reach thermal equilibrium with the radiation it is absorbing \cite{Aros:2008ef}. The heat capacity of the black hole horizon is displayed for some particular values of $d$ and $n$ in figures (\ref{figuraCEH}) for the EH solution, (\ref{figuraCPL}) for the Pure Lovelock solution and (\ref{figuraCBHS}) for solution with the $n$-fold degenerated GS solution. In this figures one can observe the existence of a phase transition. This occurs at $r_+=r_*$, where the temperature of black hole horizon reaches a local maximum. Moreover, at the left side of the phase transition, for $r_+<r_*$, the heat capacity is always positive, whereas at the right side the heat capacity is negative. 

For values of the $M$ parameter $M \in ]M_{cri1},M_{cri2}[$, where there are three horizons, we observe that, for each ordered pair $(r_+,r_{++})$, such that $r_+>r_*$, $C_+<0$ and $C_{++}>0$. Conversely, for $r_{+}<r^*$, the local heat capacity becomes $C_+ >0$. This is due to the phase transition mentioned above. The evolution will be discussed in the next section. 

 \section{Evolution of three horizon solution}
 
It is worth to stress that the local definitions of heat capacity $K$ and $C$ share the same behavior. Both have changes of sign and phase transitions in the same points, due that both share the same dependence on the factor $(dT/da)^{-1}$. See Eqs.(\ref{C1aley1},\ref{C1aley2},\ref{C1aley3},\ref{C1aley4}) for $C$ and Eqs.(\ref{PureLovelockKexplicit},\ref{nFoldHeat}) for $K$. 

The range of interest corresponds to $M_{cri1} < M <M_{cri2}$ where there are three horizons. The first thing to notice is that $T_{++}$ is an always increasing function of $r_{++}$ for $M_{cri1} < M < M_{cri2}$. Moreover, $K_{++}>0$ and $C_{++}>0$. This implies that the cosmological horizon always {\it aims} to reach thermal equilibrium with the radiation that absorbs. 

The situation of the black hole horizon is quite different. First, there is an addition value of the mass parameter $M^{*} \in [M_{cri1},M_{cri2}]$  that defines a phase transition where $K_+$ and $C_+$ go from negative for $M^{*}<M<M_{cri2}$, to positive for $M_{cri1} < M < M^{*}$. This can be observed in temperature's profile, as shown above, where $T_+$ vanishes for $M_{cri1}$ then increases, as a function of $r_+$ until it reaches a local maximum when $M=M^*$ and then $T_+$ becomes a decreasing function of $r_{+}$ until it vanishes when $M=M_{cri2}$. 

There are two situations to discuss. For the $n$-fold Lovelock gravity $T_+ > T_{++}$ strictly for $M \in [M^*, M_{cri2}]$. However, Pure Lovelock solution, for instance with $d=8$ and $n=3$, the situation is the opposite and the temperature of the cosmological horizon is larger than the black hole horizon one for a range of the mass parameter. This determines two different evolutions depending of the original value of $M$. This also happens for vacuum solutions studied elsewhere.   

\subsection{The case of \textbf{$T_{+} > T_{++}$} for $M \in ]M_{eq},M_{cri2}[$} \label{BattleHorse}

This case comprehends every solution of the $n-$fold Lovelock gravity, and part of the Pure Lovelock solutions. 

\begin{itemize}
  \item \textbf{Above the phase transition:} In region of $M^* < M < M_{cri2}$ let us consider the case where the temperatures satisfy $T_{++} < T_+$ \footnote{For the vacuum solution  $T_{++} < T_+$ occurs in general, but the presence of the anisotropic fluid requires of a numerical analysis making quite complicate to check it in general}. Recall that that $K_{+} < 0$ and $K_{++} > 0$. This implies that the cosmological horizon increases its temperature as is absorbing hotter radiation coming from the black hole horizon. The black hole horizon also further increases its temperature as it is absorbing, but with a $K_+<0$, cooler radiation from the cosmological horizon. The only consistent scenario is that the mass parameter $M$ decreases. This can directly translate, see Eqs.(\ref{RelacionHorizontes1},\ref{RelacionHorizontes22}) into a decrease of $r_+$ and an increase of $r_{++}$. 
  
  \item \textbf{Below the phase transition:} For $M$ just below $M^{*}$ the temperatures satisfy $T_{++} < T_+$, but now both $K_{+} > 0$ and $K_{++} > 0$. As previously, the cosmological horizon increases its temperature as it is absorbing hotter radiation coming from the black hole horizon. Now, due to $K_+ > 0$, the black hole horizon also aims to thermalize with the cooler radiation coming form the cosmological horizon and thus it must reduce its temperature. Since the mass parameter $M$ is unique (for both horizons) the only consistent scenario is that $M$ must further decrease. As previously, this implies that a further decrease of $r_+$ and a further increase of $r_{++}$.  This process is expected to continue until both horizons reach the same temperature or $M=M_{eq}$.
\end{itemize}

\begin{figure}
\centering
{\includegraphics[width=75mm]{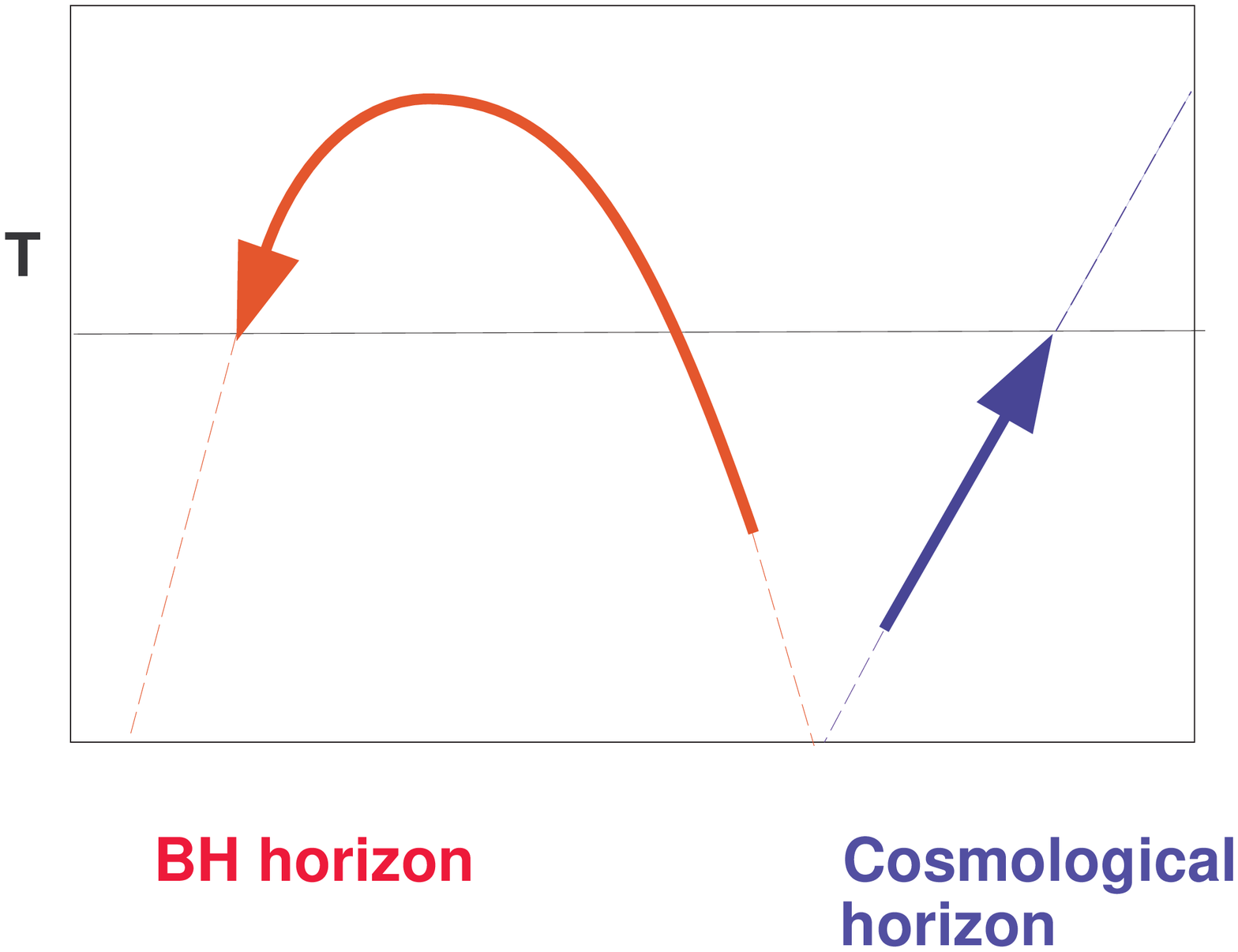}}
\caption{Evolution of both horizons.}
\label{figura1}
\end{figure}

The evolution process can be sketched from figure (\ref{figura1}). 

\begin{figure}
\centering
{\includegraphics[width=75mm]{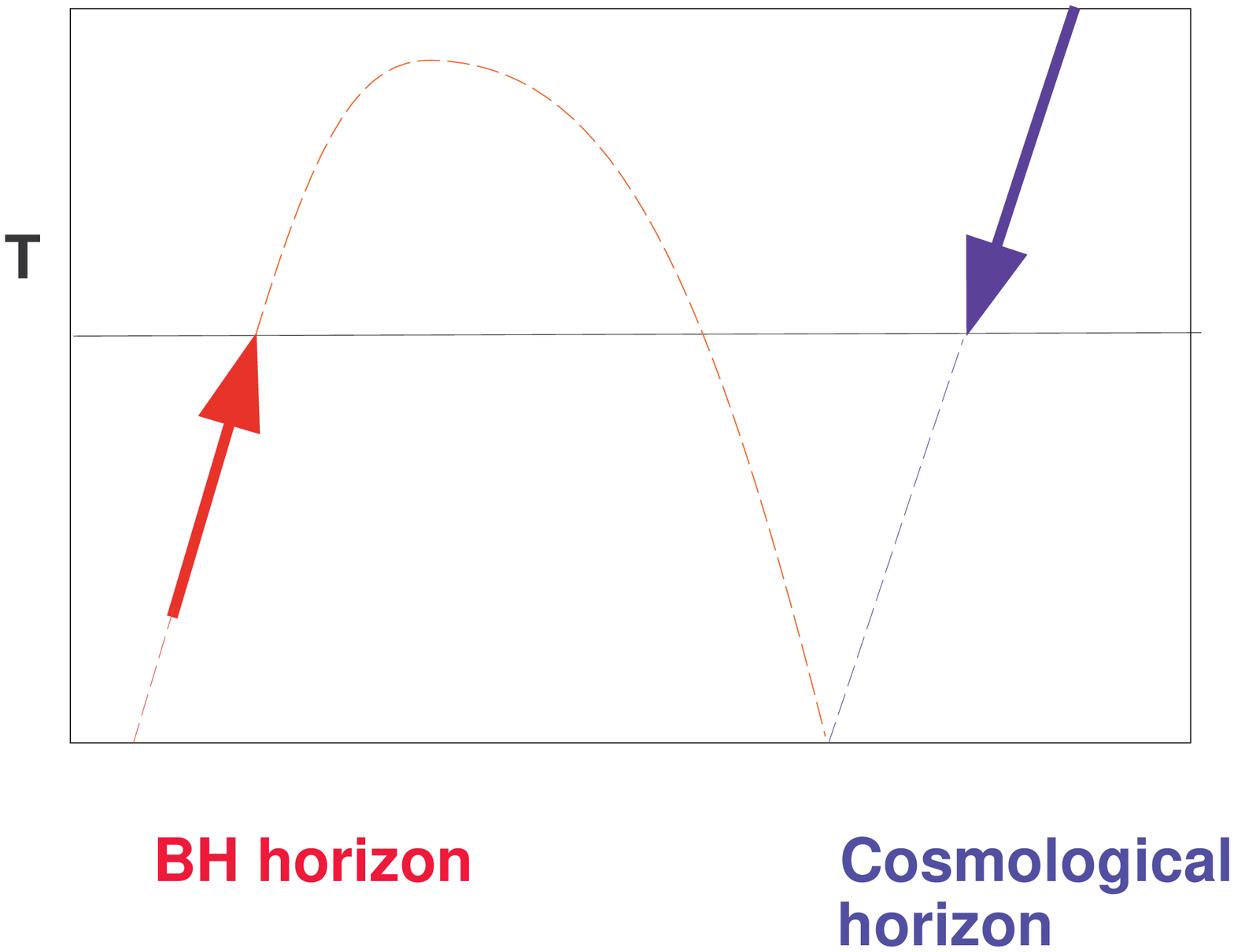}}
\caption{Evolution of both horizons.}
\label{figura2}
\end{figure}

If instead of $M>M_{eq}$ the starting point would have been a point where $M_{cri1}<M< M_{eq}$, or equivalently $T_{+} < T_{++}$, then the evolution must be in the opposite direction. This is due to $K_{+} > 0$ and $K_{++} > 0$ and therefore the black hole horizon must increase its temperature simultaneously with a decrease of the temperature of the cosmological horizon. In this case $M$ must increase its value until thermal equilibrium between both horizons is reached. This implies that the black hole horizon radius $r_+$ increases while the cosmological horizon radius decreases of $r_{++}$. This evolution has be sketched in figure (\ref{figura2}). 

The previous line of thought determines that any three horizon regular black hole must evolves until both black and cosmological horizons reach thermal equilibrium. This pictures is a complete different scenario compared with non-regular black hole geometries, such as Schw-dS, which evolve into a de Sitter space, the ground state \cite{Aros:2008ef}.   

\subsection{The case of \textbf{$T_{+} <  T_{++}$} for $M \in ]M_{CR},M_{cri2}[$ and   \textbf{$T_{+} >  T_{++}$} for $M \in ]M_{eq},M_{CR}[$}

$T_{+} <  T_{++}$ can only  happen for some of the Pure Lovelock solutions mentioned above whose mass parameter satisfies $M \in ]M_{CR},M_{cri2}[$, where $M_{CR}$ is defined by $M_{cri1}< M^*< M_{CR}<M_{cri2}$. Now, although this a almost singular behavior still it is very relevant due to its consequences in the evolution.    

\begin{figure}
\centering
{\includegraphics[width=75mm]{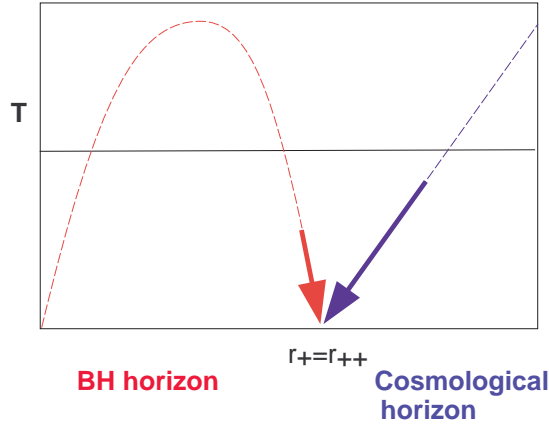}}
\caption{Evolution of both horizons.}
\label{figura3}
\end{figure}

In this case the evolution is complete different than the previous case. The heat capacity of the black horizon is always negative and the heat capacity of the cosmological horizon is always positive. As the black hole radiation has a lower temperature than the temperature of the cosmological horizon, this later must aim to reduce its temperature. Conversely, as the temperature of the black hole horizon is lower than the temperature of the cosmological horizon, then the black hole horizon is absorbing radiation and aims to reduce its temperature due to its negative heat capacity. In this way, both horizon aims to reduce their temperatures. The only compatible scenario is the mass parameter must increase and therefore the geometry must evolve into the merging of the black hole and cosmological horizons. Beyond the merging point the evolution cannot be studied by thermodynamics. This behavior was sketched in figure (\ref{figura3}).   

On the other hand, if $M \in ]M_{cr1}, M_{CR}[$ the evolution is into the thermal equilibrium  at $M=M_{eq}$. The analysis is analogous to the previous subsection (\ref{BattleHorse}). 

It is remarkable feature that in this case $M_{CR}$ actually defines a transition in the evolution of the system from evolving into thermal equilibrium between the horizon or into a extreme black hole geometry. Moreover, it is also remarkable that the mass value $M_{CR}$ doesn't define a phase transition in the (local) thermodynamics of any of the horizons, but define a property of the system/geometry as a whole.   

\subsection{The case of \textbf{$T_{+} <  T_{++}$} for $M \in [M_{cri1},M_{cri2}[$}

It is quite remarkably that for the Pure Lovelock solutions it is possible to have the conditions under which the temperature of the black hole horizon is always lower than cosmological horizon one, $T_{++} > T_{+} \forall M \in [M_{cri1},M_{cri2}[$ .

\begin{figure}
\centering
{\includegraphics[width=75mm]{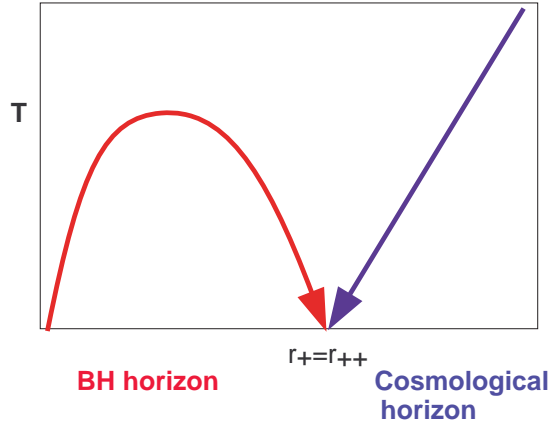}}
\caption{Evolution of both horizons.}
\label{figura4}
\end{figure}

In this case let us consider a starting point in the range $M \in ]M_{cri1},M^*]$. The case $M \in ]M^*,M_{cr2}[$ is contained in this case as will be shown. The first thing to notice is that the cosmological horizon decreases its temperature, since $K_{++}>0$, and thus, the mass parameter must increases. By other hand, the black hole horizons has specific heat $K_{+}>0$, and thus, increases its temperature up to its maximum value in $M=M^*$. Next, for the range in  $M \in [M^*,M_{cri2}]$, the cosmological horizon continue to decrease its temperature due to $K_{++}>0$, and the black hole horizon also aims to further reduce its temperature due to $K_+<0$. Thus, both horizons reduce their temperatures until both reach $T_+ = T_{++} =0$, or equivalently $r_{+}=r_{++}$ and $M=M_{cri2}$. This process has be sketched in figure (\ref{figura4}). 

\section{Conclusions and Discussion}

In this work it has been studied the problem of black solution in the presence of a anisotropic fluid that mimics the some of quantum effect that could prevent the formation of singularities. This is done for two different Lovelock theories that share to have a single de Sitter ground state. Similar to the vacuum scenario these solution present a cosmological and black hole horizons. However, the presence of the fluid introduces a third, and inner, horizon. These solutions can have two and one horizons as well, being the two solution is what can usually refer to as an extreme or zero temperature solution. There is always at least one horizon, but this can happen in two very different scenarios. In fact, it is shown that there is  a range of the mass parameter of solution $M \in [M_{cr1},M_{cr2}]$ where the solution present three horizons.    

The thermodynamics of these solutions is explored by a local method that provides with a local first law of thermodynamics at each horizon. In this respect in this work is shown that the local method can be implemented for Lovelock theories in the presence of anisotropic fluids. Although, the more traditional, and non-local, thermodynamic relations cannot be obtained directly, in this work some non-local relations are obtained as well.  

Analysis of the solutions by means of the local and non local thermodynamic relations mentioned above share a similar thermodynamic behavior. This, on the other hand, precludes that any solution must evolve until black and cosmological horizons reach thermal equilibrium or in some particular cases evolve into an extreme (black hole) geometry. The first scenario differs from the vacuum solutions, which under a similar analysis, evolve into the ground state, a de Sitter space. The evolution into the extreme black hole geometry is shared by the vacuum solutions in the corresponding cases.      

We provide a new toy model of regular black hole, whose energy density is a $d$ dimensional generalization of \cite{Dymnikova1}. This one confirms the results just mentioned explicitly.

\begin{acknowledgments}
RA would like to thank UNAB for some financial support.

\end{acknowledgments}
\bibliography{DeSitterbib}
\end{document}